\documentclass{aa}
\usepackage{graphicx,psfig}
\begin{document}
   \title{Star formation in gaseous galaxy halos}

   \subtitle{VLT-spectroscopy of extraplanar \ion{H}{ii}-regions in 
\object{NGC\,55}\thanks{Based on observations 
collected at the European Southern Observatory, Paranal (Chile); Proposal 
No.: 64.N-0399(A,B)}}

\author{R. T\"ullmann\inst{1}, M. R. Rosa\inst{2}\fnmsep\thanks{Affiliated to the 
Space Telescope Operation Division, Research and Scientific Support Department of 
the European Space Agency}, T. Elwert\inst{1}, D. J. Bomans\inst{1}, A. M. N. 
Ferguson\inst{3}, 
   \and R.-J. Dettmar\inst{1}
         }

   \offprints{R. T\"ullmann,\\
   \email{tullmann@astro.rub.de}
   }

\authorrunning{T\"ullmann et al.}

\institute{Astronomisches Institut, Ruhr-Universit\"at Bochum, Universit\"atsstr. 150, 
D-44780 Bochum, Germany
          \and
Space Telescope Coordinating Facility, c/o 
European Southern Observatory, Karl-Schwarzschild-Str. 2, D-85748 Garching, Germany
          \and
Max-Planck Institut f\"ur Astrophysik, Karl-Schwarzschild-Str. 1, D-85748 Garching, Germany
             }

   \date{Received 4 March 2003 / Accepted 22 September 2003}

\abstract{We present first deep VLT-spectra of a new class of compact 
extraplanar objects located in the disk-halo interface of the edge-on spiral 
galaxy NGC\,55. Their spectra reveal continuum emission from stars and show typical emission-lines as observed in ordinary disk \ion{H}{ii}-regions. 
With respect to emission-line fluxes, these spectra are very similar to those 
obtained for the diffuse ionized gas (DIG), except [\ion{O}{iii}]$\lambda 5007$ which is strongly decreased by more than a factor of 3.
Similar to the DIG the prominent ionization stage of oxygen is O$^{+}$, 
whereas the corresponding one for low metallicity \ion{H}{ii}-regions is O$^{++}$.
A comparison with CLOUDY model simulations reveals that the ionization mechanism of these compact objects is most likely photoionization by late OB stars (O9.5 to B0). Further analysis of diagnostic diagrams unambiguously confirms the \ion{H}{ii}-region character. 
This raises the question whether these extraplanar \ion{H}{ii}-regions (EHRs) originated from the prominent extraplanar gas of this galaxy or have just been expelled from the disk into the halo.  
From hydrodynamical considerations ejection from the disk can be ruled out. 
Therefore, these objects must have formed within the halo. Compared to the average abundance of the central disk \ion{H}{ii}-region ($45\%\,Z_{\odot}$) both EHRs reveal substantially lower [O/H] abundances of about $10\%\,Z_{\odot}$. 
We could establish for the first time strong differences in the metal content along the minor axis of this galaxy. Oxygen appears to be less abundant in the halo by about a factor of 4.   
Since both EHRs are located above the central part of NGC\,55, it appears likely, 
that their formation was triggered by star formation activity in the disk below. 
In this environment the molecular gas clouds out of which EHRs have formed can 
survive and collapse only in the period between two successive bursts of star formation.
  
}
\maketitle
%
\section{Introduction}
The formation of massive stars in galaxies of different Hubble Types
has been studied comprehensively during the last decades. For spirals
this process is considered to be almost exclusively concentrated to
the disk (Kennicutt \cite{kenni}, Pilyugin \cite{pily}). On the other
hand, the existence of a young OB-star population in the halo of the
Milky Way galaxy (e.g., Greenstein \& Sargent \cite{green}, Kilkenny
et al. \cite{kilke}), as well as in external galaxies (e.g., Comer\'on et 
al. \cite{comer1}, Comer\'on et al. \cite{comer2}) has been well 
established. 
In order to clarify the origin of this young stellar component different 
scenarios have been proposed: ejections from the disk as a consequence of 
supernova (SN) induced runaways (Stone \cite{stone}); ejections from stellar
clusters as a result of gravitational encounters (Leonard \& Duncan
\cite{leo2}, Conlon \cite{conlon}); and star formation in the halo itself 
(Keenan \cite{keenan},
Hambly et al. \cite{ham}, Ferguson \cite{fergi2}, Ferguson et
al. \cite{fergi3}). For a review see Tobin (\cite{tobin}) and
references therein.

The subject of the present paper is a new class of gaseous, fairly
compact, and isolated objects which are located in the halo of several
edge-on galaxies. Such extraplanar ionized regions are visible on
H$\alpha$-imaging data for NGC\,55 (Ferguson et al. \cite{fergi}),
\object{NGC\,891} (Howk \& Savage \cite{howk}), \object{NGC\,3628}
(Fabbiano et al. \cite{fabbi}), \object{NGC\,4244} (Walterbos
\cite{walterbos}), \object{NGC\,4388} (Gerhard et al. \cite{gerhard}),
\object{NGC\,4402} (Cortese et al. \cite{cortese}), \object{NGC\,4634} (Rossa 
\& Dettmar \cite{rossa}), and \object{NGC\,5775} (T\"ullmann et al. 
\cite{me}).  Typical $|z|$-heights (distance above the disk) range from 
0.7\,kpc in NGC\,891 to about 3.0\,kpc in NGC\,5775. 

From a morphological point of view, these objects appear very much
like a small scale disk \ion{H}{ii}-region with embedded
clusters of massive stars formed recently. If the \ion{H}{ii}-region
character of these extraplanar objects can be confirmed, new observational 
light would be shed on the origin of the general halo OB-star
population.  Moreover, this might also constrain the origin of runaway
halo stars that must have been formed in the halo and that are
expected to be accompanied by faint diffuse gas clouds (Keenan
\cite{keenan}).

Finally, the hypothesis of extraplanar star formation (ESF) has to be
verified against alternative scenarios mentioned above, namely ejection from 
the disk or collision between cloudlets within high velocity clouds (Dyson \& Hartquist \cite{dyson}). 

In order to investigate the possibility of ESF, we took the first spectra of two prominent regions located in the disk-halo interface of NGC\,55. This SBm galaxy is a member of the Sculptor Group at a distance of $D=1.6$\,Mpc and is known to possess several clumps of ionized gas lying in the disk-halo interface (Graham \& Lawrie \cite{graham}, Ferguson et al. \cite{fergi}, and Otte \& Dettmar \cite{otte}). As it is seen nearly edge-on ($i=80\degr$), disk and halo are well separated.

For NGC\,55, ejection of complete objects (gas and ionizing stars)
from the disk can be ruled out by timing arguments for velocities
below 130 km\,s$^{-1}$.  Up to this speed of interstellar cloud
material the flight duration in $z$-direction to reach 1 kpc is much
larger than the typical lifetime of the ionizing source.  
Furthermore, even a small amount of interstellar matter sitting along the path
would lead to a separation of cloud and embedded stars because of the
enormous difference in impact parameters cloud vs. cloud as compared
to stars vs. cloud.

Knowledge of the gas phase abundances could help to distinguish
between different creation mechanisms of the extraplanar ionized
regions. Rather low metallicities compared to the disk abundances
would indicate that these regions have formed from almost pristine
local halo material. Relatively high abundances would restrict the
origin of the clouds to material processed in star-forming regions of
the disk. Furthermore, abundance determinations of other components of
the interstellar medium (ISM), such as the diffuse ionized gas (DIG,
Dettmar \cite{dettmar}) or classical \ion{H}{ii}-regions would also
help to constrain the origin of the gaseous halo. \ion{H}{ii}-regions
in the disk, preferentially in the center, are of particular
importance, because they provide reliable reference abundances of the
processed gas. Once chemical abundances have been constrained, the
assumed transport of gas from the disk into the halo and back (Shapiro
\& Fields \cite{shapi}, Norman \& Ikeuchi \cite{norman}) can be probed
directly.

Therefore, the intention of this paper is threefold: first to
determine the nature of the gaseous objects in the halo; second, to derive
abundances and compare them to those measured in the disk and the disk-halo interface;
and third, to discuss their possible origin.

A description of observational strategy and data reduction is given in
Sect.~2. We then discuss the results from our first VLT-based
optical multi-object-spectroscopy (MOS) and H$\alpha$-imaging data of
the investigated extraplanar objects, the DIG, and the central
\ion{H}{ii}-region of NGC\,55 (Sect.~3). Finally, Sect.~4 
provides essential conclusions and briefly summarizes the most important
results of this work.

\section{Observations and data reduction}
\begin{figure*}
\centering
\includegraphics[angle=0,width=11.7cm,height=11.2cm]{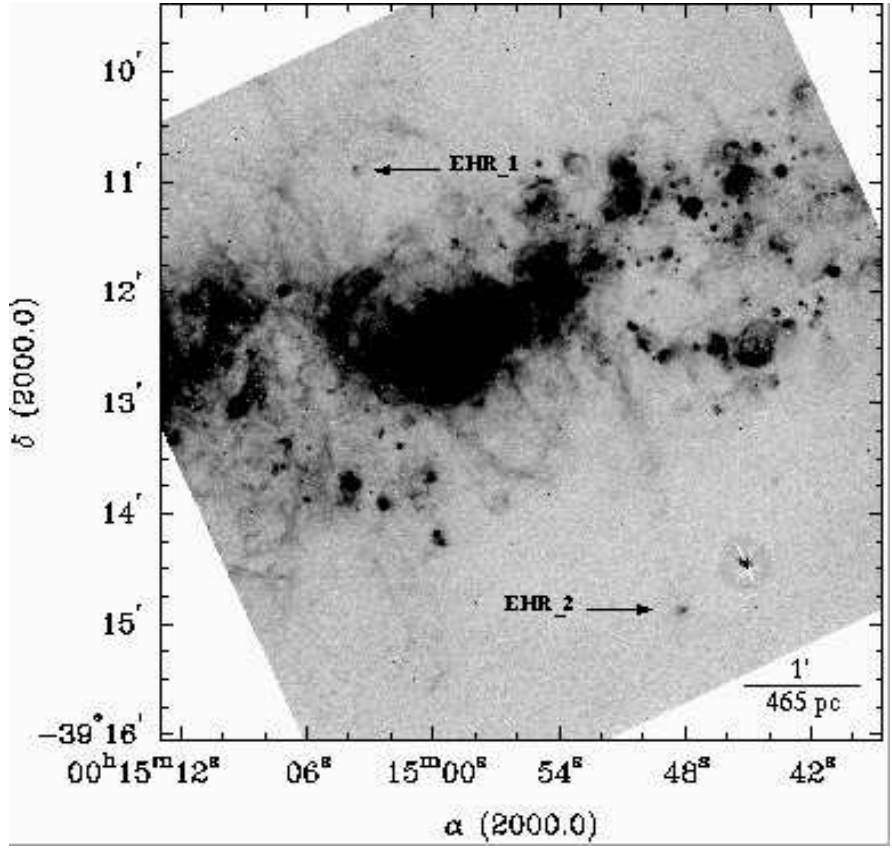}
\caption{The central region of NGC\,55. This continuum-subtracted H$\alpha$-image has 
been obtained with UT1$+$FORS1 at the VLT. Various spectacular DIG-features, such 
as filaments, superbubbles and colliding shells are well resolved. Extraplanar 
\ion{H}{ii}-region candidates are marked by arrows. The total integration time amounts 
to 20\,min.}
\label{fig1}
\end{figure*}

The dataset presented here has been obtained at ESO Paranal Observatory with VLT UT1 and 
FORS1 in imaging and MOS-mode during the nights of 9 -- 11 December 1999.
Both instrument configurations were used in standard resolution, giving a 
field of view of $6\farcm 8 \times 6\farcm 8$ and a pixel scale of $0 \farcs 2$ 
pix$^{-1}$.   
A Tektronix $2048 \times 2046$ CCD detector with a pixel size of $24 \times 24\ \mu$m 
was used for all measurements.

\subsection{H$\alpha$ pre-imaging}
For optimal placement of the MOS slitlets, pre-imaging of the target field was 
required. Therefore, a 20\,min exposure of the central part of NGC\,55 has been taken with FORS1 in H$\alpha$ (H\_Alpha$+$59). In order to remove the stellar continuum from the H$\alpha$-frame, an additional 2\,min exposure has been obtained in the R-band using the R\_BESSEL$+$36 filter.
During the integration, seeing conditions were slightly below $1\farcs 0$.

Data reduction was carried out in the usual manner using standard IRAF software (V2.11.3). 
Cosmic ray hits were removed from the single exposure using the Laplacian edge 
detection algorithm (van Dokkum \cite{dokkum}) rather than filtering and threshold depending methods.  

\begin{figure*}
\centering
\includegraphics[angle=0,width=6.5cm,height=6cm]{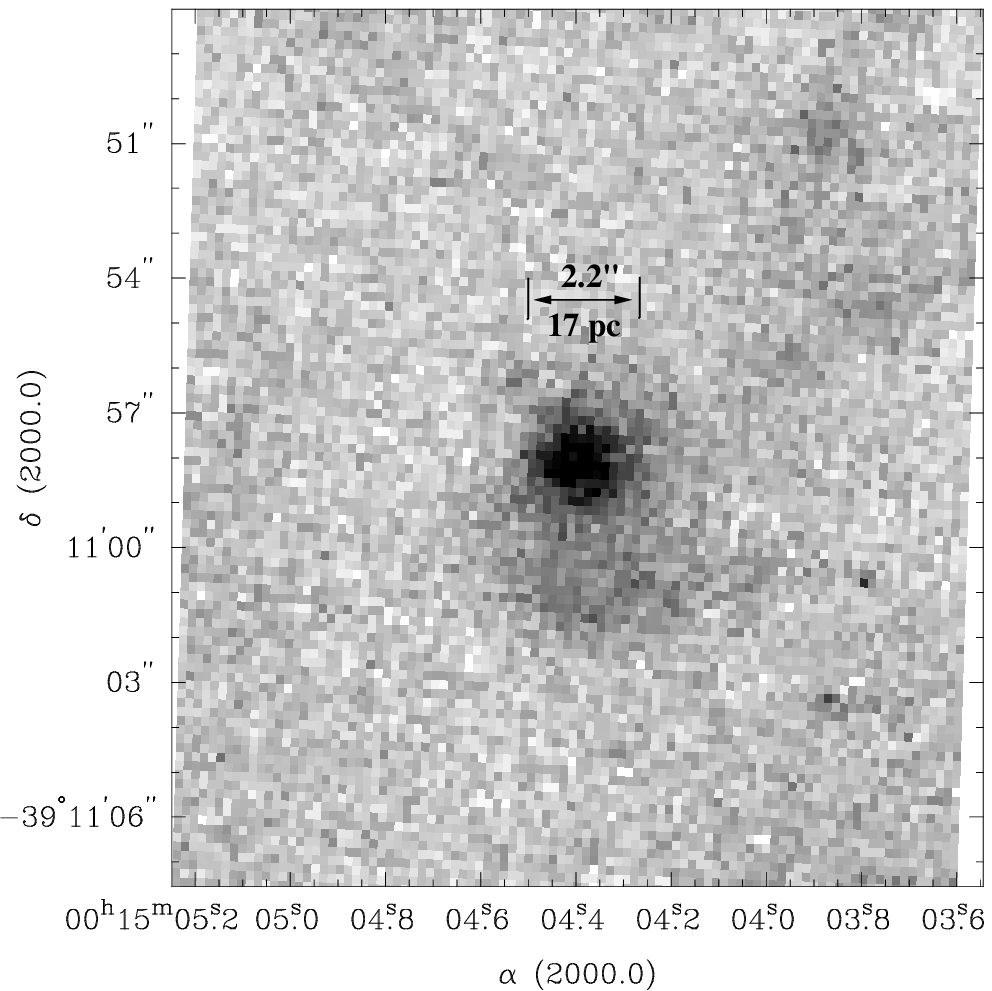}
\hspace{.5cm}
\includegraphics[angle=0,width=6.5cm,height=6cm]{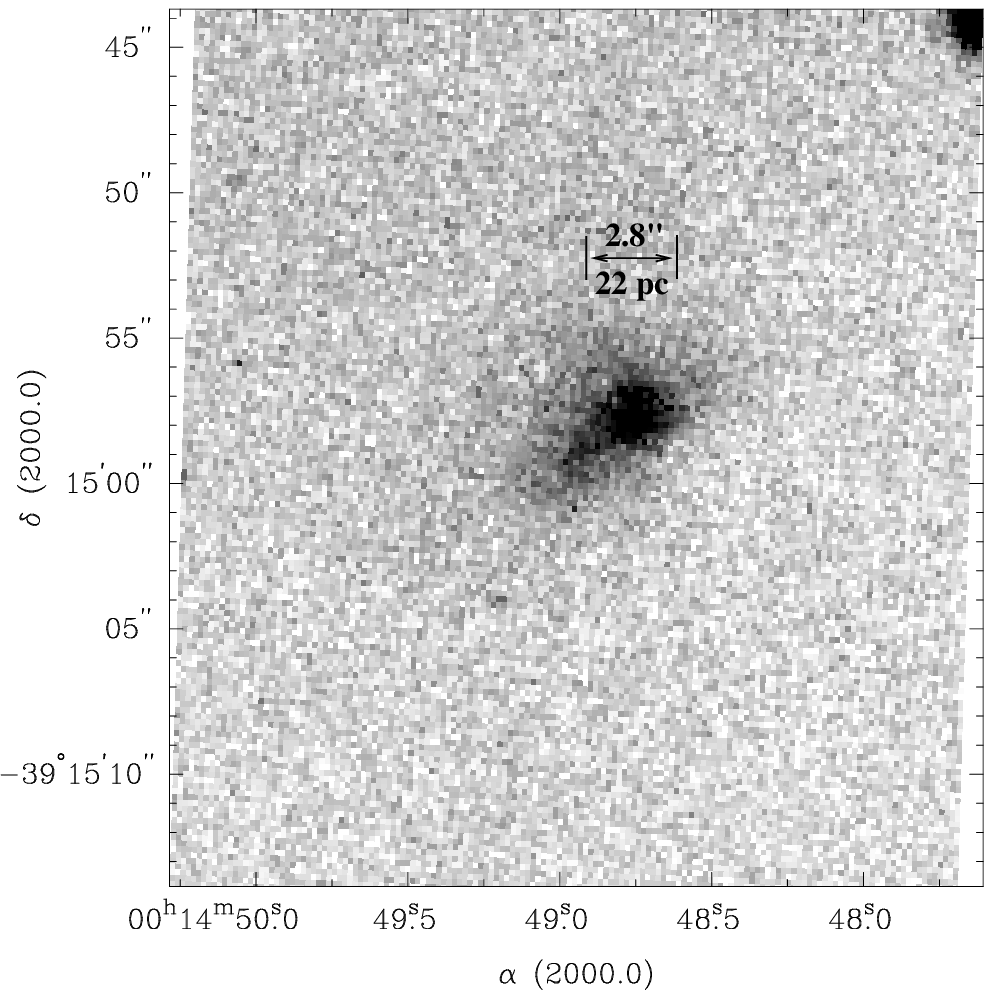}
\caption{Blow-up views of EHR\_1 in the northern (left) and EHR\_2 in the southern halo 
(right).
Both objects show a dense central core with diffuse H$\alpha$-emission in their outskirts which is sharply bounded towards the halo. The coordinates (J2000) for EHR\_1 are: R.A.: 
$00^{\rm h} 15^{\rm m}\,04\fs 387$, Dec.: $-39\degr 
10\arcmin 58\farcs130$ and for EHR\_2: R.A.: 
$00^{\rm h} 14^{\rm m}\,48\fs 737$, Dec.: $-39\degr 
14\arcmin 57\farcs840$.}
\label{fig2}
\end{figure*}

The final processed and continuum-subtracted H$\alpha$-image is shown in Fig.~\ref{fig1}.
Magnified versions of the two EHR candidates are presented in Fig.~\ref{fig2}. 
It is remarkable, that in both cases the diffuse emission appears to be truncated towards the halo of NGC\,55. If confirmed for more EHR objects in other galaxies as well and supported by kinematical data this would give evidence for the interaction of dense cloud material with the rarefied gas in the halo.

\begin{figure*}
\centering
\vspace{1cm}
\hspace{2.cm}
\includegraphics[angle=0,width=12.5cm,height=11cm]{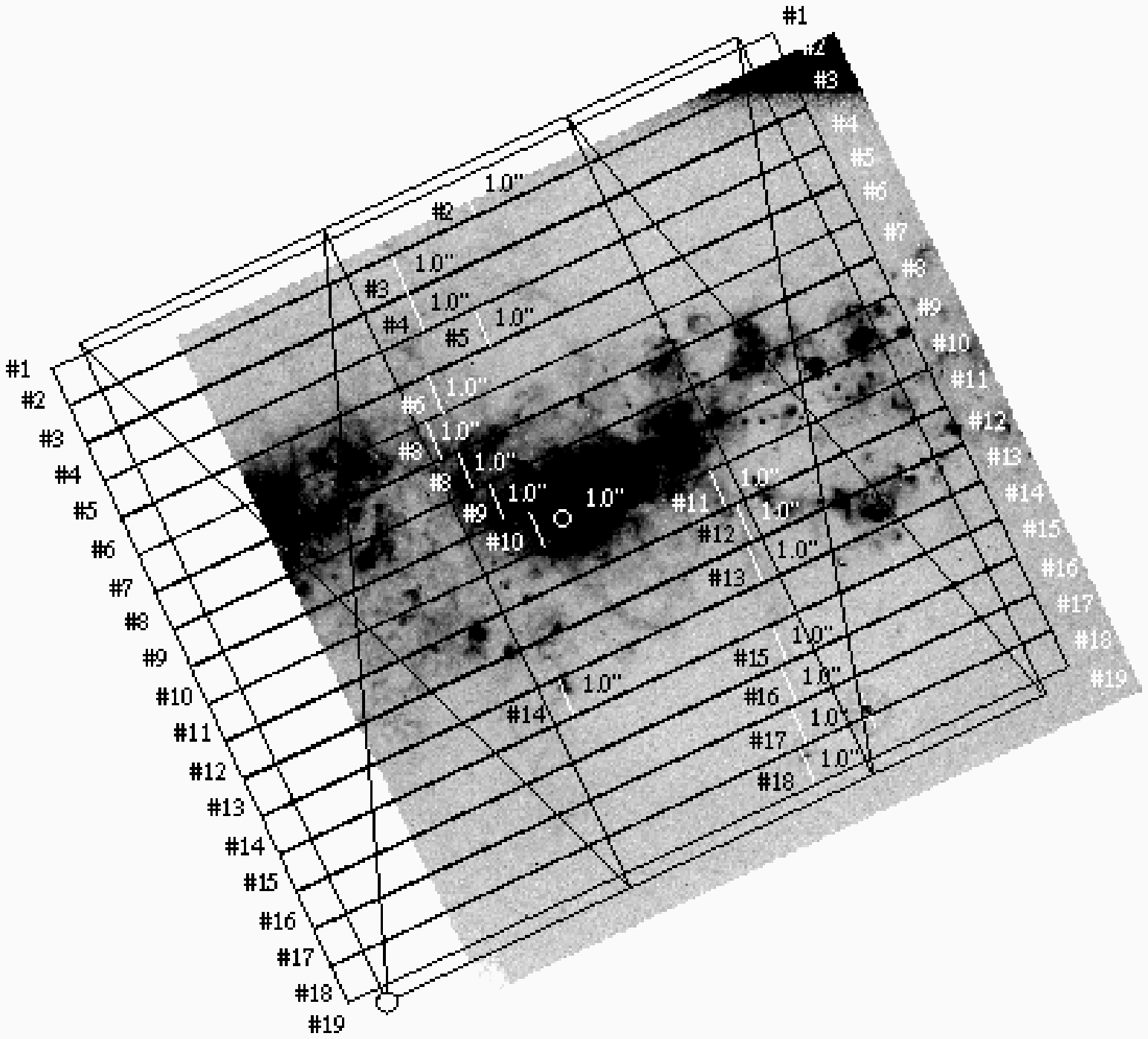}
\vspace{-.3cm}
\caption{The FORS1 MOS-mask with chosen slit positions. Slitlets
\#5 and \#18 cover EHR\_1 and EHR\_2, while slitlets 
\#10 and \#13 cut through the central \ion{H}{ii}-region and the DIG, respectively.  
Slit \#2 has been allocated for nightsky subtraction. The 
pointing coordinate (J2000) is: R.A.: 
$00^{\rm h} 14^{\rm m}\,57\fs 579$ and Dec.: $-39\degr 11\arcmin 58\farcs53$.}
\label{fig3}
\end{figure*}

\subsection{Multi-object-spectroscopy}
For spectroscopic observations, two grisms were selected which measure most of 
the important diagnostic emission-lines in the optical.
GRIS\_600B$+$12 covers the blue wavelength range from 3500\,\AA\ -- 5900\,\AA, while GRIS\_600R$+$14 was used between 5200\,\AA\ and 7400\,\AA. 
However, the effective wavelength coverage depends mainly on the position of the slitlet on the CCD. Fig.~\ref{fig3} displays the FORS1 MOS-mask and the individual positions of the 19 slits. Only slitlets within the unmarked part of the CCD chip will produce spectra within the tabulated wavelength range.

Each slitlet has a constant length of $22\arcsec$ and matches typical length 
scales within EHRs or the DIG very well.
A slit width of $1\farcs 0$ was chosen for all slitlets which yields a spectral resolution of about 5.4\,\AA\ for the red and 6.0\,\AA\ for the blue grism, respectively.
The morphology of the gaseous structures put some limitations on the orientation of the MOS-mask and observations could not be carried out along the parallactic angle. 
Since FORS1 at UT1 is equipped with a longitudinal atmospheric dispersion corrector (Avila et al. \cite{avila}), this effect could be partially compensated for.

Seeing conditions during spectroscopic observations improved from $1\farcs 5$ at the beginning to $0\farcs 9$ at the end of the night.
For each grism a total integration time of 1.5\,hrs was spent on-source, split into 2 exposures of 45 min. each. 

The spectroscopic data were reduced with ESO MIDAS software (01SEPpl1.3, 
context long) following standard procedures. 
All spectra of scientific interest (slitlets \#5, 18, 10, 13, and 2, cf. Fig.~\ref{fig3}) and their corresponding calibration frames were copied to individual files and were then treated as normal longslit frames. 

In order to obtain the pure emission from the central \ion{H}{ii}-region (slitlet \#10) as well as from both EHRs (slitlets \#5 and \#18), nightsky and background emission close above and below these regions was averaged and subtracted. 
For the DIG (slitlet \#13), which is completely free of continuum emission, nightsky removal was performed by subtracting the nightsky spectrum obtained with slitlet \#2.

Finally, measured emission-line fluxes were normalized to H$\beta$ and were dereddened using the Balmer lines of H$\alpha$, H$\beta$, and H$\gamma$, adopting the standard interstellar extinction curve by Osterbrock (\cite{oster}), and assuming his case B approximation with an electron temperature of $T=10^4$\,K.  Differences between theoretical and dereddened decrements of ${\rm H}\alpha/{\rm H}\beta$ and ${\rm H}\gamma/{\rm H}\beta$ are less than 3\%, hence uncertainties in the reddening correction are negligible.  Since galactic foreground extinction is insignificant towards the direction of NGC\,55 (Burstein \& Heiles \cite{burstein}), this is mainly a correction for extinction internal to NGC\,55.
Effects of hydrogen absorption in the underlying stellar continua on the derived reddening are not detectable.

\begin{figure*}
\centering
\includegraphics[width=5.1cm,height=8.8cm,angle=-90,clip=t]{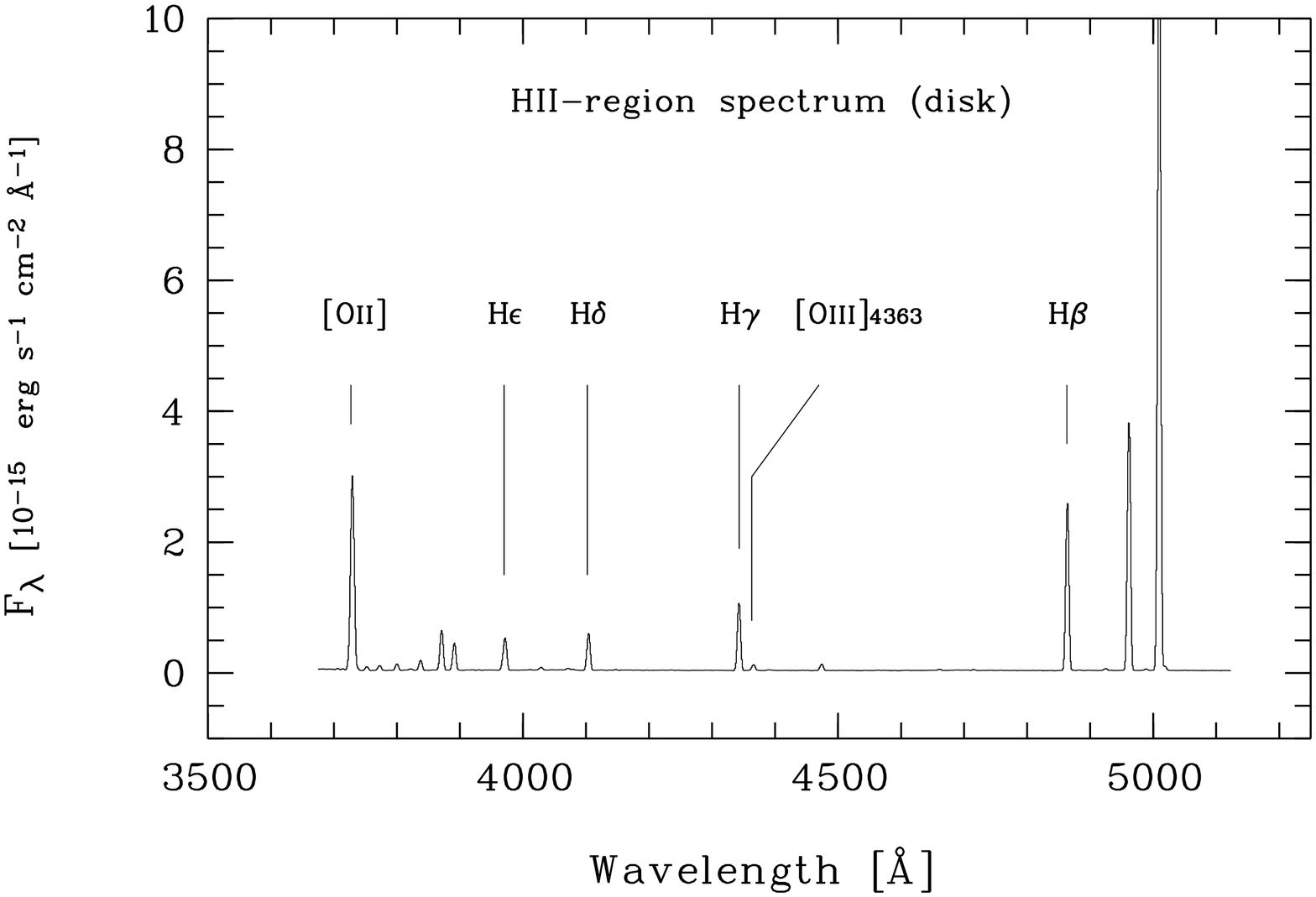}
\hfill
\includegraphics[width=5.1cm,height=8.8cm,angle=-90,clip=t]{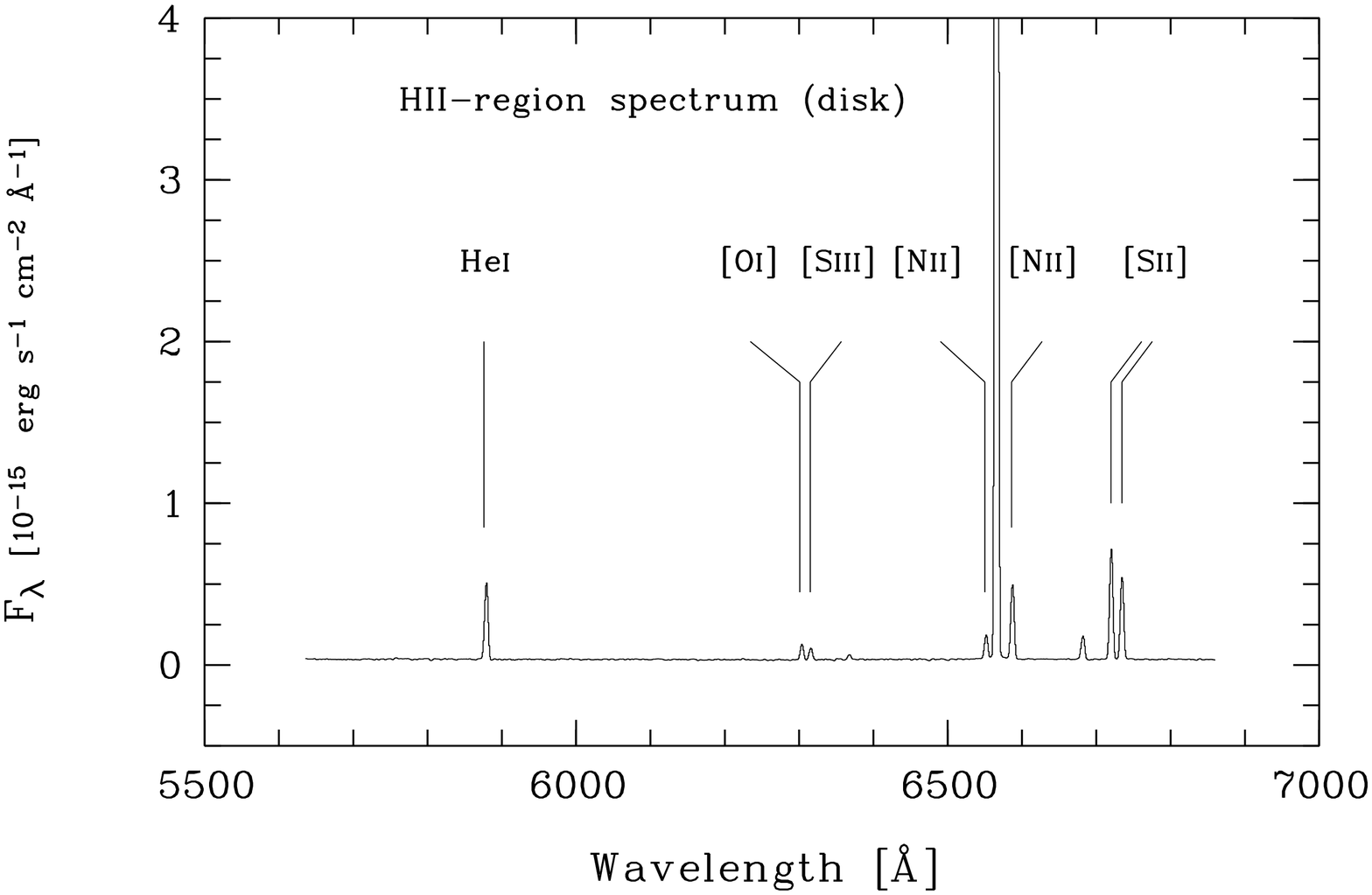}
\includegraphics[width=5.1cm,height=8.8cm,angle=-90,clip=t]{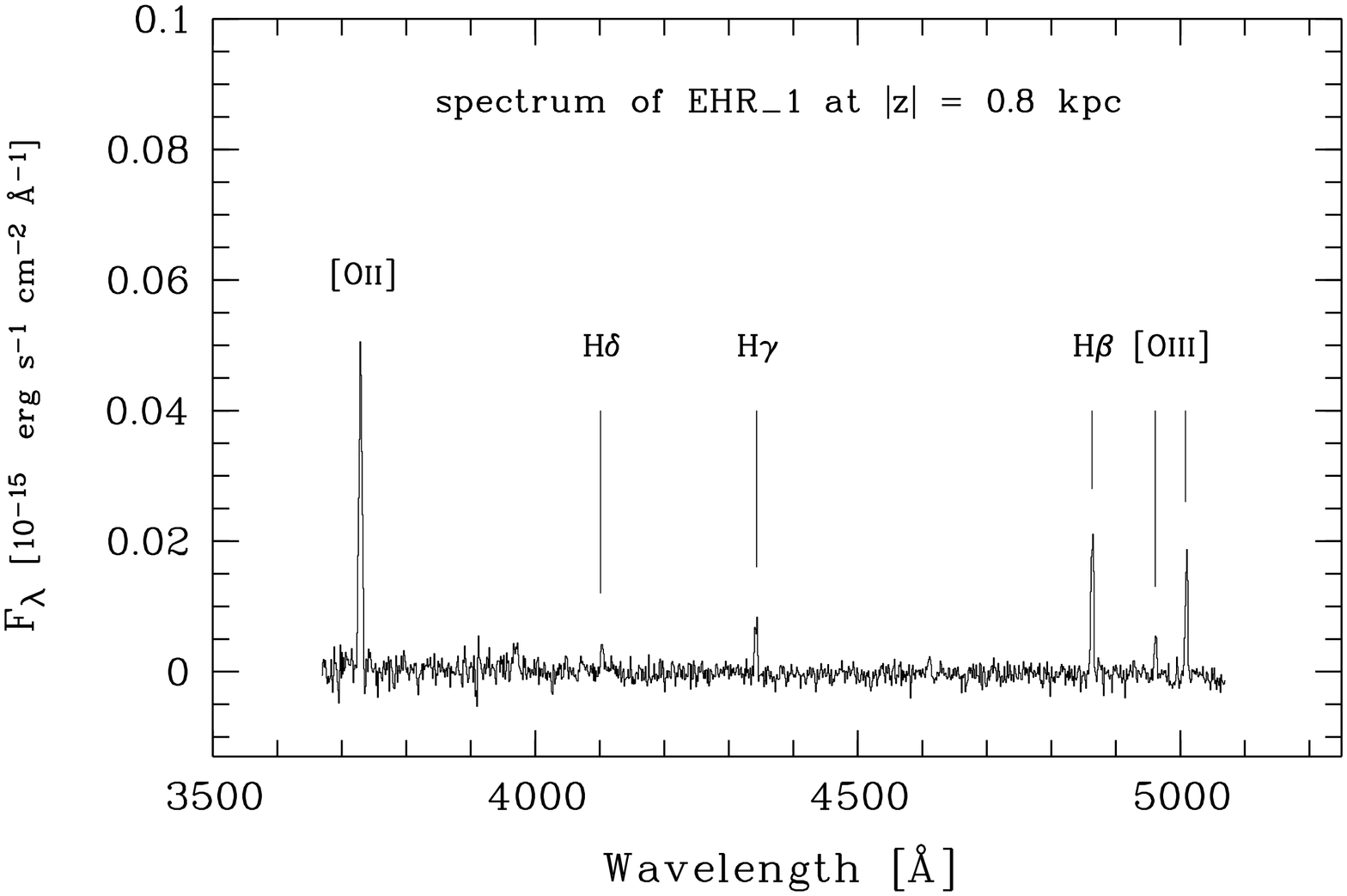}
\hfill
\includegraphics[width=5.1cm,height=8.8cm,angle=-90,clip=t]{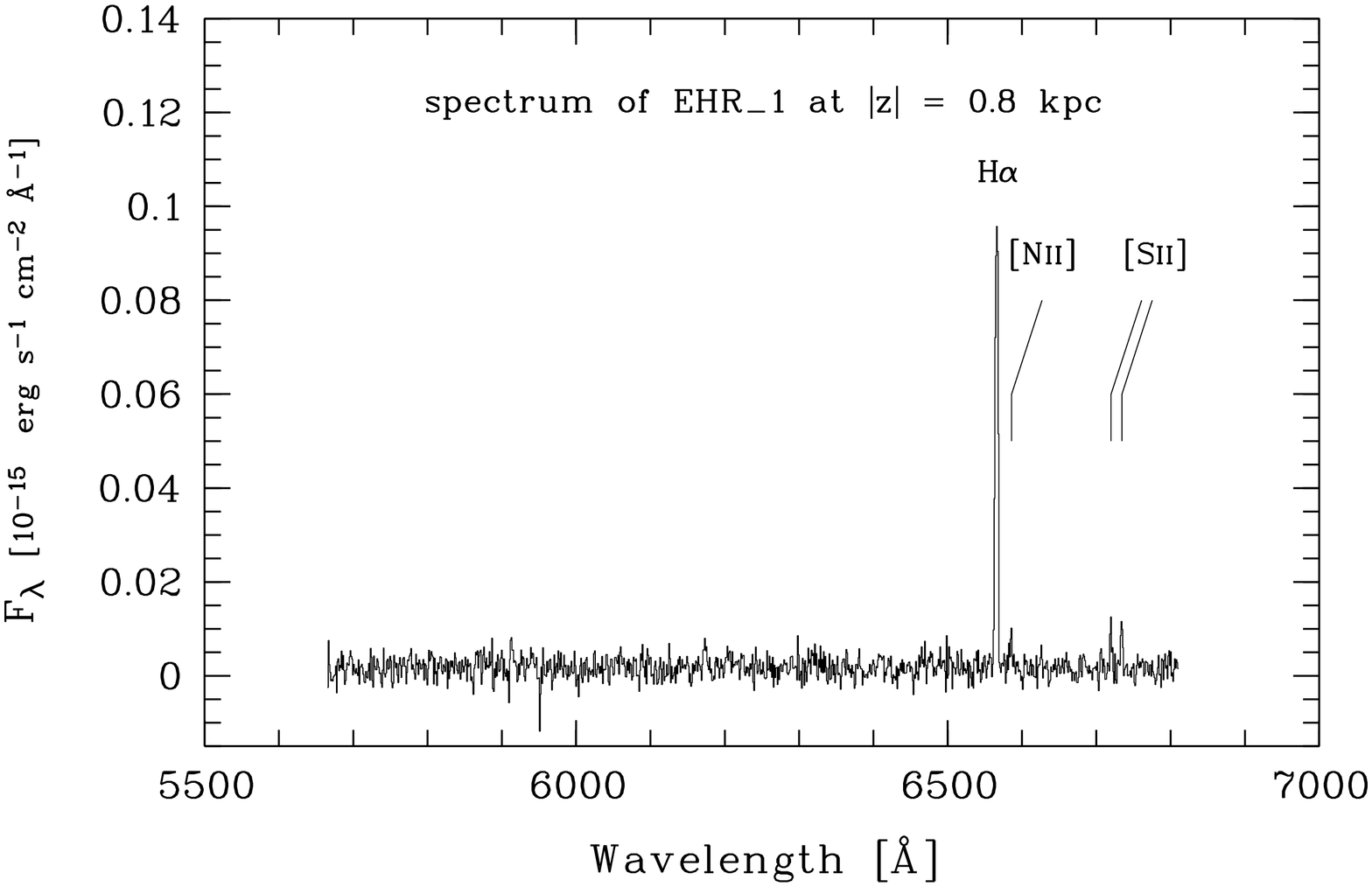}
\includegraphics[width=5.1cm,height=8.8cm,angle=-90,clip=t]{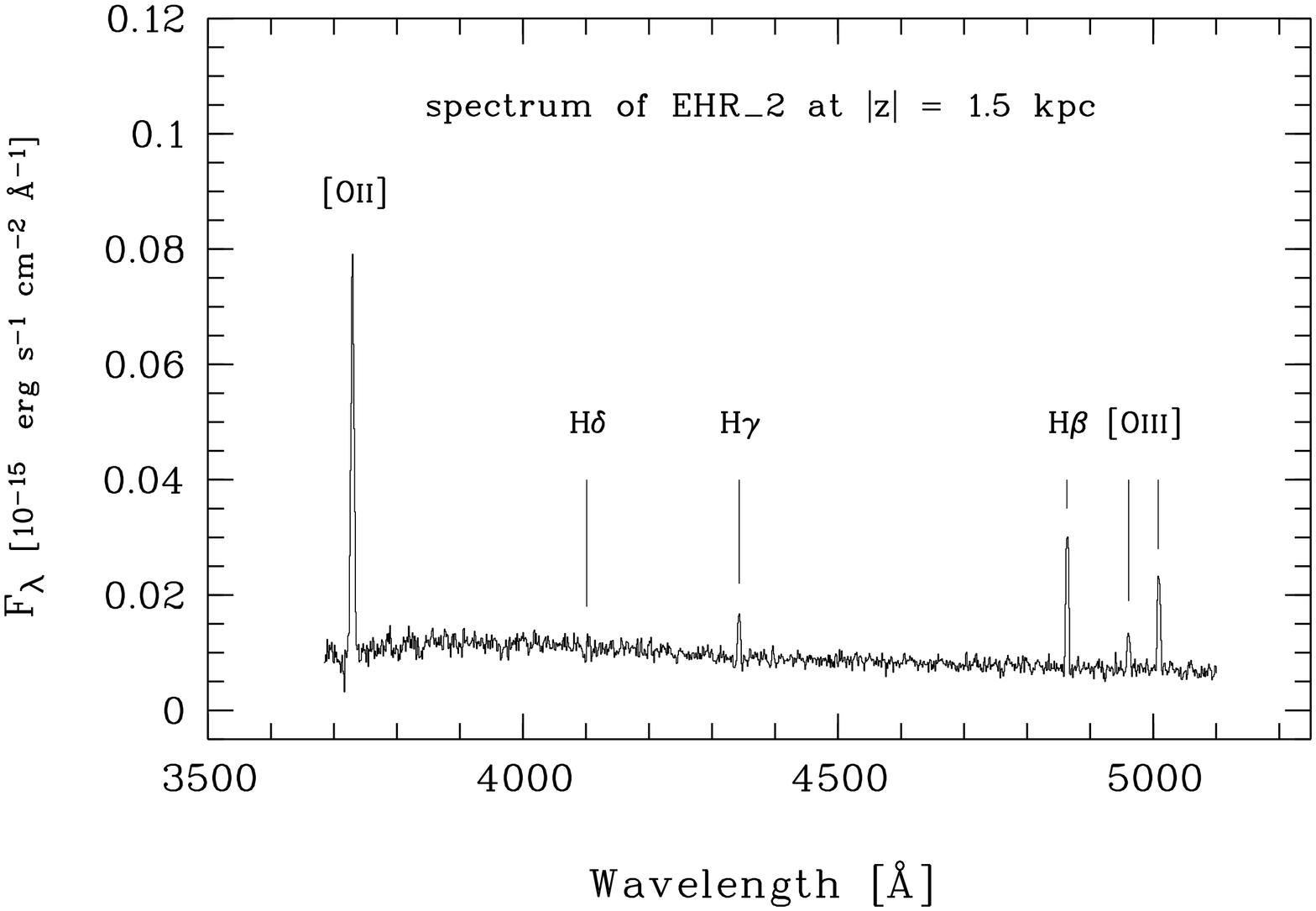}
\hfill
\includegraphics[width=5.1cm,height=8.8cm,angle=-90,clip=t]{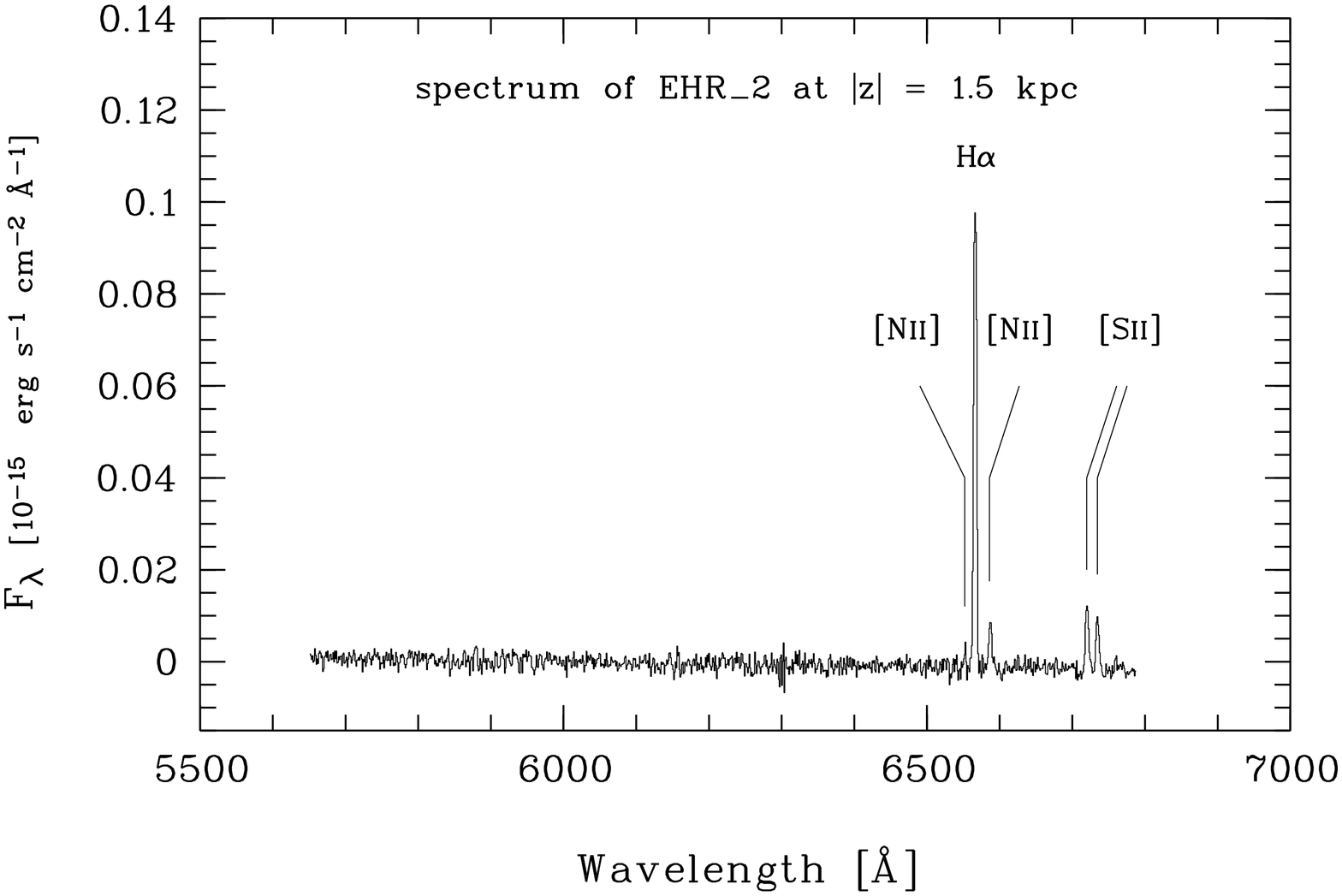}
\includegraphics[width=5.1cm,height=8.8cm,angle=-90,clip=t]{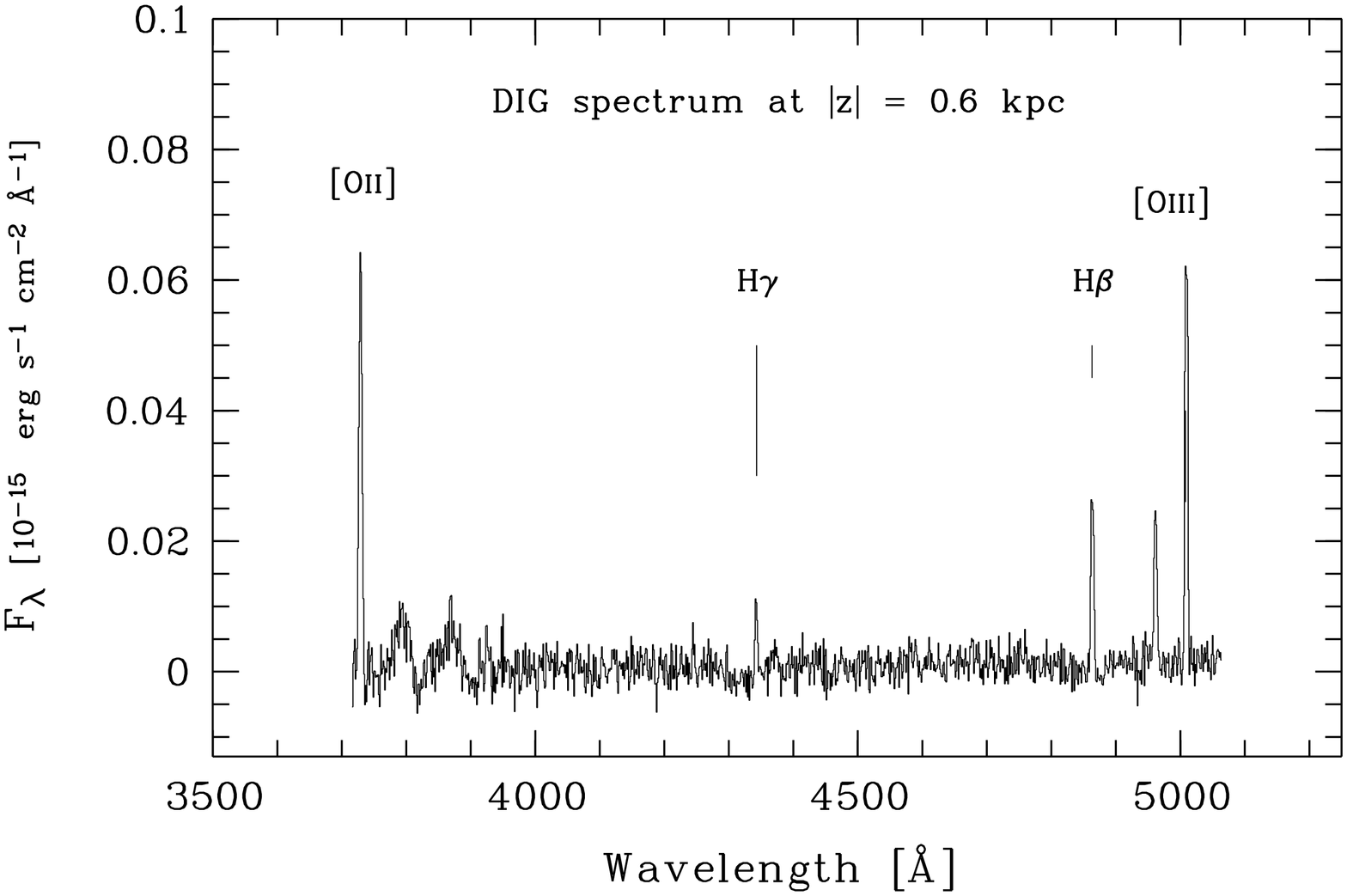}
\hfill
\includegraphics[width=5.1cm,height=8.8cm,angle=-90,clip=t]{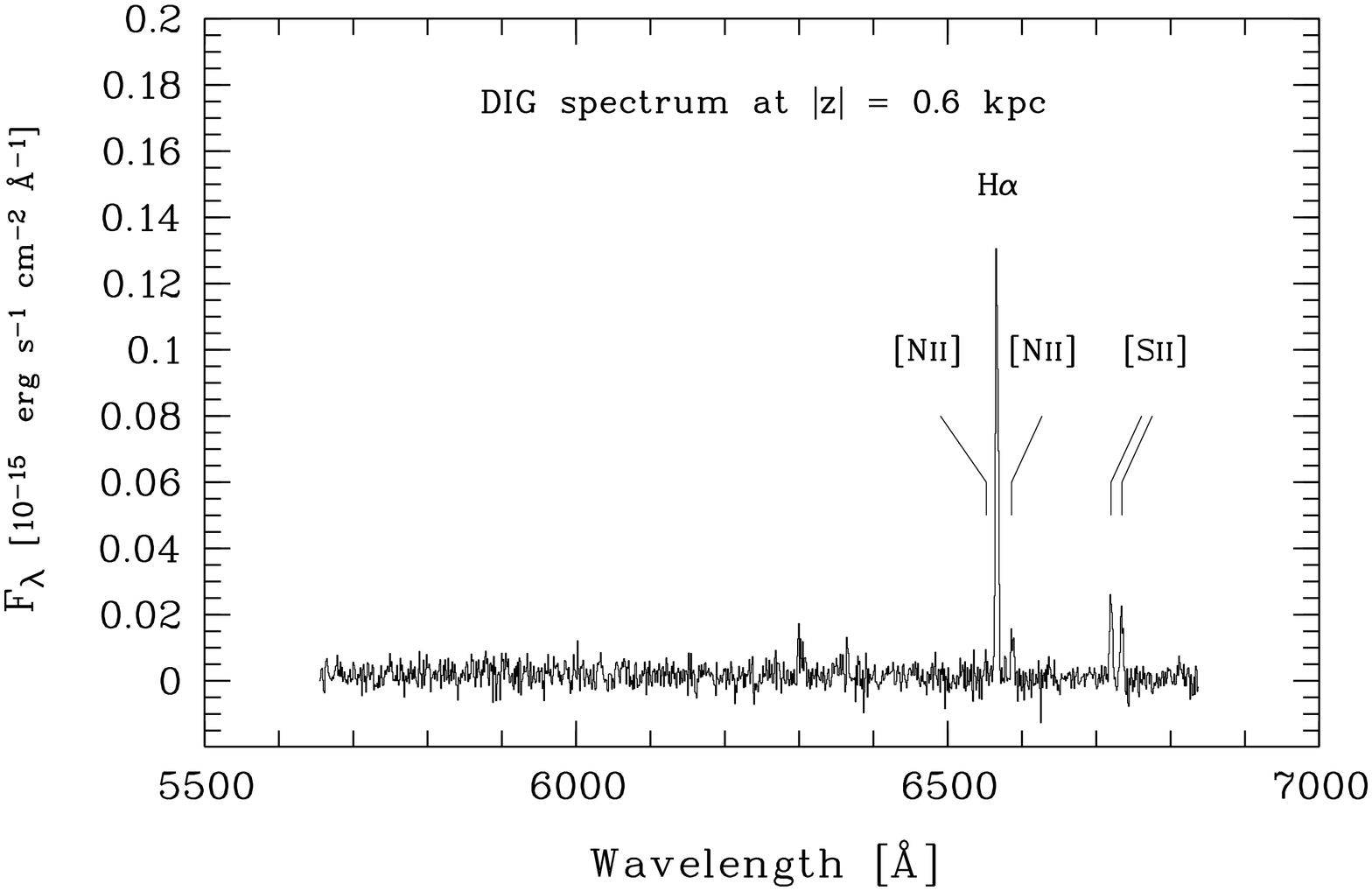}
\caption{Integrated spectra of a classical \ion{H}{ii}-region (uppermost panel), the two EHRs, and the DIG (lowermost panel).  Although the DIG has a similar spectrum compared to the EHRs, it appears to be ionized by a slightly harder radiation field, as implied by a significantly higher flux in the [\ion{O}{iii}]$\lambda$5007 emission-line.}
\label{fig4}
\end{figure*}
\section{Results and discussion}
\subsection{The nature of the extraplanar objects}
The H$\alpha$-image presented in Fig.~\ref{fig1} provides up to now the most detailed 
and spatially best resolved view of the central part of NGC\,55. 
A huge curved filament of gas and dust, anchored in the disk, is protruding out of the image 
plane, apparently pointing towards R.A.\,(2000): 
$00^{\rm h} 15^{\rm m}\,07^{\rm s}$ and Dec.\,(2000): $-39\degr 12\arcmin 00\arcsec$. 

\begin{figure*}
\centering
\includegraphics[width=6.5cm,height=8.7cm,angle=-90,clip=t]{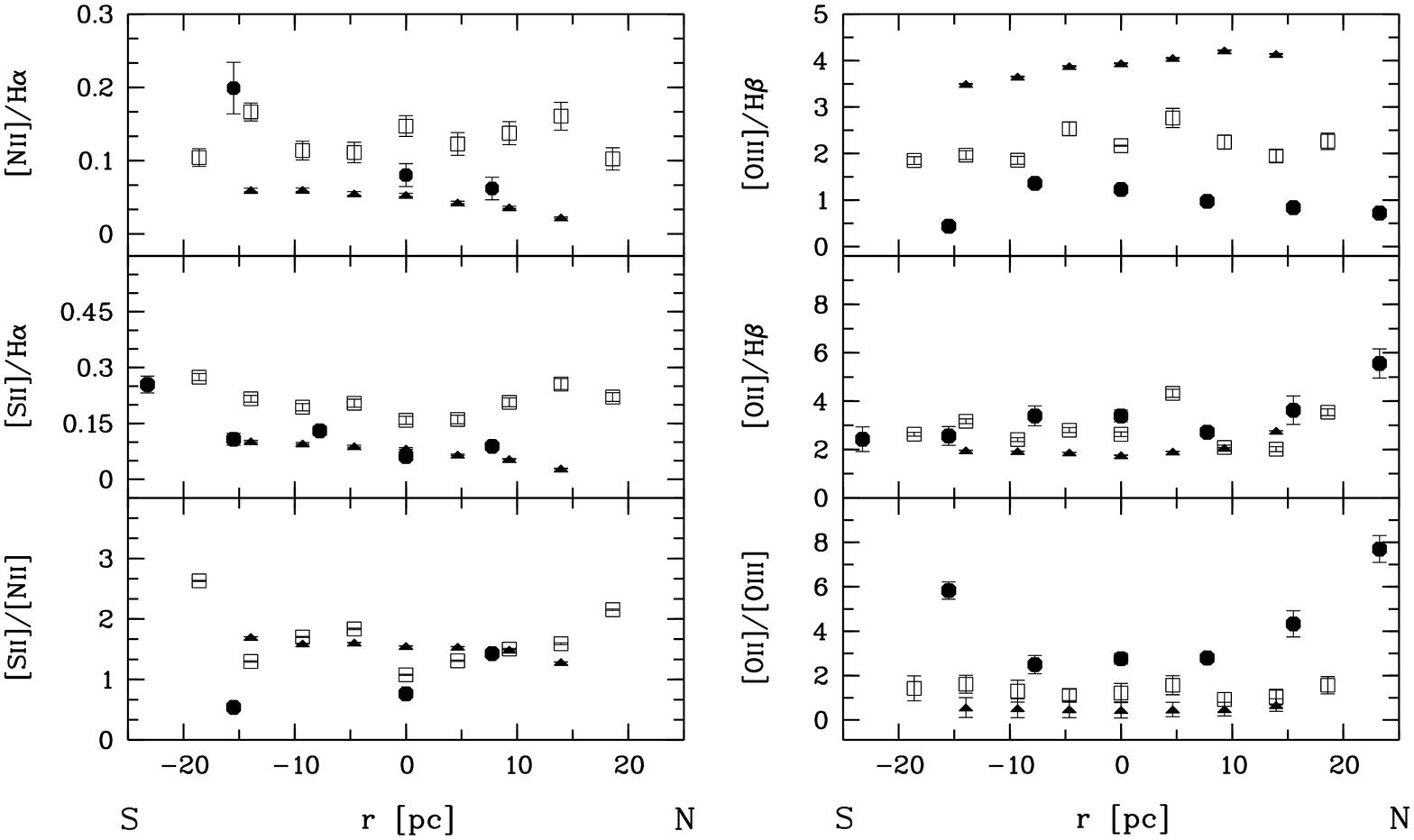}
\hspace{0.4cm}
\includegraphics[width=6.5cm,height=8.7cm,angle=-90,clip=t]{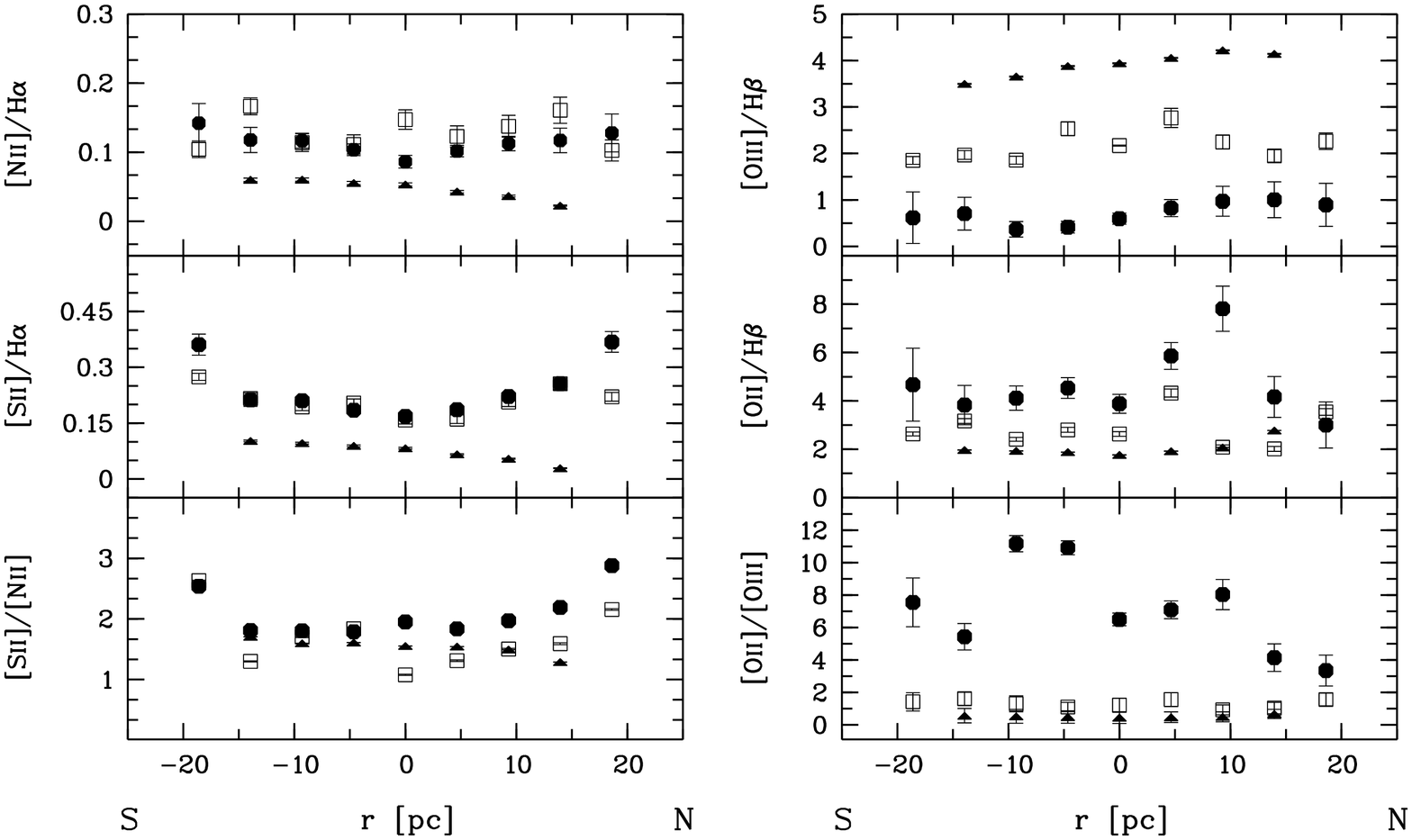}
\caption{Reddening-corrected emission-line ratios for EHR\_1 (left panels) and EHR\_2 (right panels) as a function of $r$, the spatial distance from their center (filled circles).
Compared to the DIG (open squares) and the classical \ion{H}{ii}-region (filled triangles) significant differences among the individual radiation fields become apparent, most clearly seen in [\ion{O}{iii}]/H$\beta$ and [\ion{O}{ii}]/[\ion{O}{iii}].}
\label{fig5}
\end{figure*}
Of particular interest are the two objects marked by arrows, whose magnifications are shown in Fig.~\ref{fig2}.  Their diameters are measured by fitting single Gaussians to the peak H$\alpha$-emission and determining the corresponding FWHM. EHR\_1 in the north spans 17\,pc and EHR\_2 in the south 22\,pc in diameter.

Assuming an inclination of $i$ = 80$\degr$ (Hummel et al. \cite{hummel};
Puche et al. \cite{puche}), these regions are located at corrected
$z$-heights of 0.77\,kpc (EHR\_1) and 1.52\,kpc (EHR\_2),
respectively.  If effects along the line of sight are negligible, EHR\_1 is located within the expanding oxygen-bright SN shell that was detected 20
years ago by Graham \& Lawrie (\cite{graham}).  
It would be interesting to learn from similar observations in other galaxies 
if these extraplanar regions predominantly occur at points where such
shells, created by OB stars and SNe, intersect and the gas is piled
up.
At least the compressed gas at R.A.(2000): $00^{\rm h} 15^{\rm m}\,07^{\rm s}$ and Dec.\,(2000): $-39\degr 11\arcmin 00\arcsec$ seems to support this idea.

EHR\_2 in the southern halo is possibly located on top of an extended
($\sim 1$\,kpc) filament.  Ferguson et al. (\cite{fergi}) have
speculated whether this structure might be a chimney that has been
produced by correlated SN activity in the underlying star-forming
regions (Norman \& Ikeuchi \cite{norman}). In this case one would
expect to detect X-ray emission from the overpressured hot ionized
medium (HIM) which is funneled through the chimney into the halo. Some support for this idea is given by the 1.5\,keV ROSAT-map presented by Dahlem et al. (\cite{dahlem}) where this emission is detectable at a level of 2.5\,$\sigma$ along a substantial part of the filament.  
Due to the presence of a prominent point source, it cannot be determined if
there is also X-ray emission in the vicinity of EHR\_2. 
It remains to
be clarified if densities of the filamentary structure also support
the chimney scenario.  Nevertheless, there is a good alignment of DIG
and HIM along the filament.

The most important immediate result from our spectra, which justifies further
investigations of the ESF scenario, concerns the detection of
spatially concentrated continuum emission, which originates within the
more diffuse body of the extraplanar objects.  The morphology of the
continuum and the nebular emission-line distribution is direct
evidence for stellar sources responsible for the excitation of these
regions.  Correspondingly, the flux-calibrated and background-subtracted spectra, integrated along their total $z$-extent
(Fig.~\ref{fig4}), are very similar to low excitation
\ion{H}{ii}-regions (e.g., McCall et al. \cite{call}, Ferguson et
al. \cite{fergi98}, Bresolin et al. \cite{breso}).  For EHR\_1 in the
northern halo, continuum emission is very weak and can hardly be seen
in the presented spectrum. However, on the flux-calibrated and extinction-corrected 2D frame it is clearly visible within the innermost 9\,pc.  Fig.~\ref{fig4} also shows the spectrum for EHR\_2 (southern halo) which reveals a continuum much more prominent than that found in EHR\_1.

Line ratios for both EHRs presented in Fig.~\ref{fig5} are plotted as a function of $r$, the spatial distance in [pc] from the center. In addition, we overplotted typical emission-line data for the DIG and the central \ion{H}{ii}-region.

\begin{table*}
\caption{Reddening-free flux ratios relative to H$\beta$. Additionally, extinction 
coefficients $c_{\rm ext}$, derived gas temperatures, electron densities, and 
heliocentric velocities for all regions under consideration are listed. Measured flux errors are less than 14$\%$, for the \ion{H}{ii}-region they are below 9$\%$. 
A ``$\le$"-sign indicates upper limits. For comparison, CLOUDY model 
predictions for both EHRs are also listed, assuming ATLAS input spectra of type 
O9.5 and B0 (Kurucz \cite{kur}), electron densities of 600\,cm$^{-3}$, and stellar metallicities of $0.1\,Z_{\odot}$, respectively. Observed heliocentric velocities agree well within errors.}
\begin{tabular}{l l l l l l l}
\hline
\hline
Line & EHR\_1 & CLOUDY & EHR\_2 & CLOUDY & DIG & \ion{H}{ii}-region  \\
H$\beta = 1$ & & O9.5, 600, 0.1 & & B0, 600, 0.1 & & \\
\hline
$\Sigma\ [\ion{O}{ii}]$\,$\lambda$7325 & -- & 0.13 & -- & 0.14 & $\le$0.13 & -- \\
$[\ion{S}{ii}]$\,$\lambda$6716 & 0.20 & 0.05 & 0.44 & 0.05 & 0.18 & 0.14 \\
$[\ion{S}{ii}]$\,$\lambda$6731 & 0.19 & 0.04 & 0.33 & 0.05 & 0.12 & 0.11 \\
$[\ion{N}{ii}]$\,$\lambda$6583 & 0.17 & 0.38 & 0.31 & 0.54 & 0.20 & 0.11 \\
$[\ion{N}{ii}]$\,$\lambda$6548 & -- & 0.13 & 0.17 & 0.18 & $\le 0.09$ & 0.04 \\
$[\ion{O}{i}]$\,$\lambda$6300 & $\le$0.07 & 0.002& $\le$0.10 & 0.002& $\le$0.07 & 0.02 \\
\ion{He}{i}\,$\lambda$5876 & $\le$0.14 & 0.07 & 0.09 & 0.04 & $\le$0.02 & 0.12  \\
$[\ion{N}{ii}]$\,$\lambda$5755 & $\le$0.06 & 0.01 & $\le$0.04 & 0.02 & $\le$0.01 & 
0.002 \\
$[\ion{O}{iii}]$\,$\lambda$5007 & 0.77 & 1.31 & 0.69 & 0.52 & 1.75 & 4.21 \\
$[\ion{O}{iii}]$\,$\lambda$4959 & 0.22 & 0.46 & 0.27 & 0.18 & 0.57 & 1.42 \\
$[\ion{O}{iii}]$\,$\lambda$4363 & $\le$0.07 & 0.03 & $\le$0.05 & 0.01 & $\le$0.04 & 0.04\\
$\Sigma\ [\ion{S}{ii}]$\,$\lambda$4072 & $\le$0.04 & 0.02 & $\le$0.09 & 0.02 & $\le$0.08 
& 0.01 \\
$\Sigma\ [\ion{O}{ii}]$\,$\lambda$3727 & 3.75 & 4.19 & 4.32 & 4.71 & 4.24 & 1.88 \\
\hline
$c_{\rm ext}$ & 0.45 & -- & 0.45 & -- & 0.45 & 0.50 \\
$T_{{\rm e}}$([\ion{N}{ii}]) [K] & -- & 14\,500 & -- & 13\,600 & $\le$17\,500 & 
11\,700 $\pm 300$ \\
$T_{{\rm e}}$([\ion{O}{iii}]) [K] & -- & 15\,300 & -- & 15\,000 & 
$\le$16\,400 & 11\,400 $\pm 300$ \\
$n_{\rm e}$ [cm$^{-3}$] & 600$\pm 90$ & 600 & 100$\pm 15$ & 600 & $\le 10$ & 300$\pm 90$ \\
$v_{{\rm hel}}$ [km s$^{-1}$] & $131\pm 9$ & -- & $151 \pm 8$ & -- & $142\pm 9$ & $151 \pm 8$ \\
\hline
\end{tabular}
\label{tab1}
\end{table*}

Line ratios of [\ion{N}{ii}]/H$\alpha$ and [\ion{S}{ii}]/H$\alpha$, EHRs tend to be DIG-like rather than \ion{H}{ii}-region-like.  If compared to traditional \ion{H}{ii}-regions, this would imply the presence of a dilute and substantially softer radiation field (Mathis \cite{mathis}, Domg\"orgen \& Mathis \cite{doma}). As can be seen from [\ion{O}{iii}]/H$\beta$ and [\ion{O}{ii}]/[\ion{O}{iii}], the ionizing radiation field of EHRs actually appears to be the softest among the three components considered here. Hence, the hottest O stars of type O5 -- O7, which are usually found within traditional disk \ion{H}{ii}-regions, can likely be excluded as ionizing sources of the EHRs.

In order to further constrain the \ion{H}{ii}-region character of
these extraplanar objects, observed line ratios need to be compared to
predictions of pure photoionization models.  Therefore, a grid of model simulations has been calculated with CLOUDY94 (Ferland \cite{fer97}, \cite{fer00}), assuming a spherical and radiation-bound geometry with volume filling factors of $f_{\rm V} = 1.0$ and 0.5 and ionization parameters $U$ in the range of 0.02 -- 0.04.  We adopted the standard dust composition (Ferland \cite{fer97}) and considered electron densities between $n_{e}=100\ \mathrm{cm}^{-3}$ and $1000\ \mathrm{cm}^{-3}$. CLOUDY metallicities for the ISM are varied between $0.01\ \mbox{and}\ 1.0\,Z_{\odot}$. ATLAS (Kurucz \cite{kur}) and WM-basic (Pauldrach et al. \cite{paul}) model atmospheres of stellar types O8.5 to B1 were chosen. Although the NLTE WM-basic code considers line blanketing and is probably the most realistic approach, the grid of available stellar input spectra seems to be too coarse to distinguish between individual stellar sub-classes. 

However, to cross-check the WM-basic predictions with those produced
by the ATLAS models, a common stellar template of type O9.5 has been
selected.  All model predictions are in reasonable agreement. For the
strong emission-line of $[\ion{O}{ii}]$\,$\lambda$3727, a deviation of
about 35$\%$ occurs, whereas the effect on weaker lines
(e.g. $[\ion{S}{ii}]$\,$\lambda\lambda$6716,6731,
$[\ion{O}{i}]$\,$\lambda$6300 or $\Sigma
[\ion{S}{ii}]$\,$\lambda$4072) is negligible. The
NLTE COSTAR-code of Schaerer et al. (\cite{schae_a}, \cite{schae_b})
appears not to be appropriate, because of an overestimation of the
stellar UV photon flux.

Our results of the best fit to the observed dereddened data of 
EHR\_1 are given in Table~\ref{tab1}. Some discrepancies are
apparent for [\ion{S}{ii}]$\lambda$6716/H$\beta$ and
[\ion{S}{ii}]$\lambda$6731/H$\beta$, which can most likely be
attributed to uncertainties in determining accurate dielectronic
recombination rate coefficients.  At densities lower than in
traditional \ion{H}{ii}-regions ($< 1000\,\mbox{cm}^{-3}$), this can
affect especially the sulfur doublet easily by about a factor of 2
(Ferland et al. \cite{fer98}).  In addition, effects of changing chemical 
abundances, a combination of different stellar sources, and
varying densities are also able to alter the artificial spectrum.
If we disregard the problems related to sulfur, the
average deviation between theory and observation amounts to 40\% which
is of sufficient accuracy to establish the
\ion{H}{ii}-region character.  In view of the relatively strong
[\ion{S}{ii}] and weak [\ion{O}{i}] and \ion{He}{i}-lines shock
ionization as an alternative ionization mechanism can be excluded.
Therefore, the adopted O9.5 spectral type seems to be the best choice
for EHR\_1. An even harder ionizing spectrum would strongly
overproduce O$^{++}$ and simultaneously lower N$^{+}$ and S$^{+}$
ratios.

Comparing the best fit for EHR\_1 with corresponding observed 
flux ratios of EHR\_2 (Table~\ref{tab1}), reveals differences in
[\ion{O}{ii}]$\lambda$3727/H$\beta$,
[\ion{O}{iii}]$\lambda$4959/H$\beta$, and
[\ion{O}{iii}]$\lambda$5007/H$\beta$. These are balanced if we consider a 
softer stellar radiation field, such as produced by B0 stars.  Due to
the lack of energetic Lyman continuum photons, the intensity of
O$^{++}$ is reduced significantly which results in an increase of
O$^{+}$ emission (cf. Table~\ref{tab1}).  The averaged difference
between theoretical and observational data for EHR\_2 is lower
than that for EHR\_1 and amounts to 29\%.

In principle each line (except those of sulfur) can be fitted by both kinds of
model atmospheres, but none of the models can consistently account for 
all emission-line ratios at once. Among uncertainties in 
determining accurate recombination coefficients, a major issue of
current photoionization codes might be that the parameter adjustment
is still too coarse. The ATLAS code runs with a fixed stellar
metallicity of 1.0\,$Z_{\odot}$ but provides sufficient coverage of
effective stellar temperatures, whereas WM-basic models switch between 
two different metallicities but provide a rather coarse
grid of stellar atmospheres ($\Delta T=5\times 10^3$\,K).

In summary, we conclude that both sets of models cover observational
line ratios within a factor of 2 and restrict the ionization
sources to be of late OB type (O9.5 to B0). This is consistent
with the direct detection of late-type luminous OB stars inside EHR\_2
from HST/WFPC2 imagery (Ferguson \cite{fergi2}, Ferguson et
al. \cite{fergi3}).  Hence, this study also supports the hypothesis of
low-excitation \ion{H}{ii}-regions.

Finally, an analysis of diagnostic diagrams (e.g., Baldwin et
al. \cite{bald}, Osterbrock \cite{oster}) revealed no indications of
shock ionization and also suggests that these objects are genuine 
\ion{H}{ii}-regions.

Since the H$\alpha$-luminosity of EHR\_2 has been determined by
Ferguson et al. (\cite{fergi}) to be $L_{{\rm H}\alpha}=3\times
10^{36}\ {\rm erg\ s}^{-1}$, we can also constrain the number of stars
within this object.

According to Osterbrock (\cite{oster}), the total luminosity of a nebula 
at a given frequency  $\nu$ can be written as: 
\begin{equation}
L_{{\rm H}\alpha}= Q(H^0,T)\ (\alpha_{{\rm H}\alpha}^{{\rm eff}}
(H^0,T)/ \alpha_{{\rm B}}(H^0,T))\ h\nu_{{\rm H}\alpha},
\label{eq1}
\end{equation}
where $Q(H^0,T)$ represents the total number of ionizing photons per time 
interval, $\alpha_{{\rm H}\alpha}^{{\rm eff}}(H^0,T)$ addresses the effective 
recombination coefficient for the specific transition, and 
$\alpha_{{\rm B}}(H^0,T)$ is the sum of all recombination coefficients above 
the ground level. With the tabulated $Q(H^0,T)$ values for B0 and O9.5 stars 
(Osterbrock \cite{oster}) and with corresponding numbers for the recombination 
coefficients Eq.~(\ref{eq1}) restricts the H$\alpha$-luminosity to the interval: 
\begin{equation}
2.1 \times 10^{36}\ {\rm erg\ s}^{-1} \le L_{{\rm H}\alpha}\le 4.0 \times 10^{36}\ {\rm erg\ s}^{-1}. 
\end{equation}

This implies that the stellar content of EHR\_2 consists most likely of a single O9.5 star or a single B0 star if we assume uncertainties in the flux measurement of about 30$\%$ .

\subsection{Determination of gas phase abundances}
We now proceed to derive temperatures, densities, and chemical abundances. To obtain the best possible abundance calibration and to reliably derive the metal content of the gas, we have made use of three methods. 

The direct method converts observed intensities into emissivities from ionic transitions and is therefore based on knowledge of the gas temperature. For brighter sources (e.g., the central \ion{H}{ii}-region) electron temperatures can be derived directly from observed emission-line strengths. If temperatures cannot be measured directly, reasonable assumptions have to be made using other constraints. Since the emissivity of collisionally excited levels is strongly dependent on temperature, the chemical abundances derived without a firm knowledge of electron temperatures are only indicative.  

For the calculation of temperatures, densities, and ionic abundances we use the nebular abundance tool (NAT) based on the 5-level-atom program of De Robertis et al. (\cite{derobert}). Conversion of ionic abundances to total chemical abundances makes use of the ICF scheme of Mathis \& Rosa (\cite{maro}).

We also estimated abundances using the empirical bright line R$_{23}$-method (Pagel et al. \cite{pagel}; Edmunds \& Pagel \cite{ed}; McGaugh \cite{mcg91}, 
\cite{mcg94}; Oey \& Shields \cite{oey}). After R$_{23}$ has been obtained, the corresponding $\log(\mathrm{O/H})$ value is determined with respect to the ionizing spectrum. For our objects here the values for $\log(\mathrm{[\ion{N}{ii}]/[\ion{O}{ii}]})$ are all significantly less than $-1$. This puts the data points below the turnover point and indicates low metallicities (McGaugh \cite{mcg91}). Chemical abundances for He and N are scaled with respect to oxygen by following Oey \& Shields (\cite{oey}).
 
The third method serves as a consistency check and is purely based on photoionization models (see Skillman \cite{skill1}). In a first iteration the photoionization model code CLOUDY was used to obtain estimates for the ionization parameter $U$, gas densities and temperatures, spectral type of the ionizing sources, and stellar metallicities. Then a grid of models was calculated for metallicities of 1.0, 2/3, 1/2, 1/4, 1/10, and 1/100 times the value of solar metallicity\footnote{We consider the metallicity $Z$ to be equal to $\log({\rm O/H})$, since oxygen is the most abundant and efficient coolant} (Skillman et al. \cite{skill}).
These models provide plots of emission-line ratios vs. electron temperature.
The region where the models intersect the observed line ratios provides an estimate for the scaling factor relative to the solar [O/H]-abundance at a given gas temperature. 
All other element abundances simply scale by this value. Since the grid of 
metallicities is rather coarse, this method is considered to yield at least reliable lower and upper limits of the metal content.

In fact, this approach is very similar to the NAT-method. The main difference is a graphical solution of the emissivity relation instead of a numerical method in NAT. Moreover, NAT implicitly accounts for ICFs through the use of a pre-selected stellar ionizing spectrum.
\begin{figure}
\centering
\includegraphics[width=5.5cm,height=8.75cm,angle=-90,clip=t]{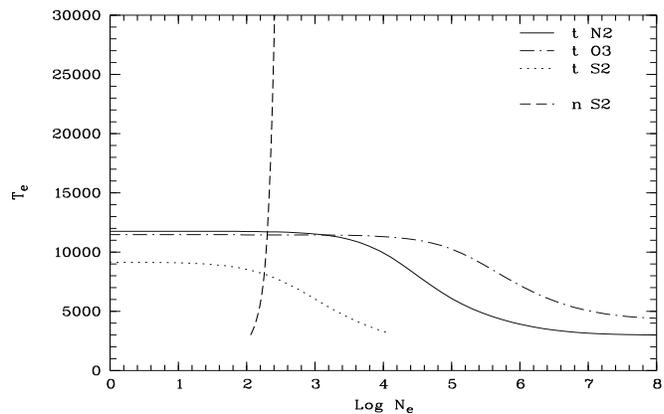}
\caption{{Plot of electron density versus electron temperature for the disk 
\ion{H}{ii}-region in NGC\,55. As can be seen S$^{+}$ emission-lines 
provide reliable temperatures for low density regions only, whereas the other two temperature indicators remain constant in density over three orders of magnitudes.}}
\label{fig7}
\end{figure}
\begin{figure*}
\centering
\hspace{0.005cm}
\includegraphics[width=5.9cm,height=6.8cm,angle=0,clip=t]{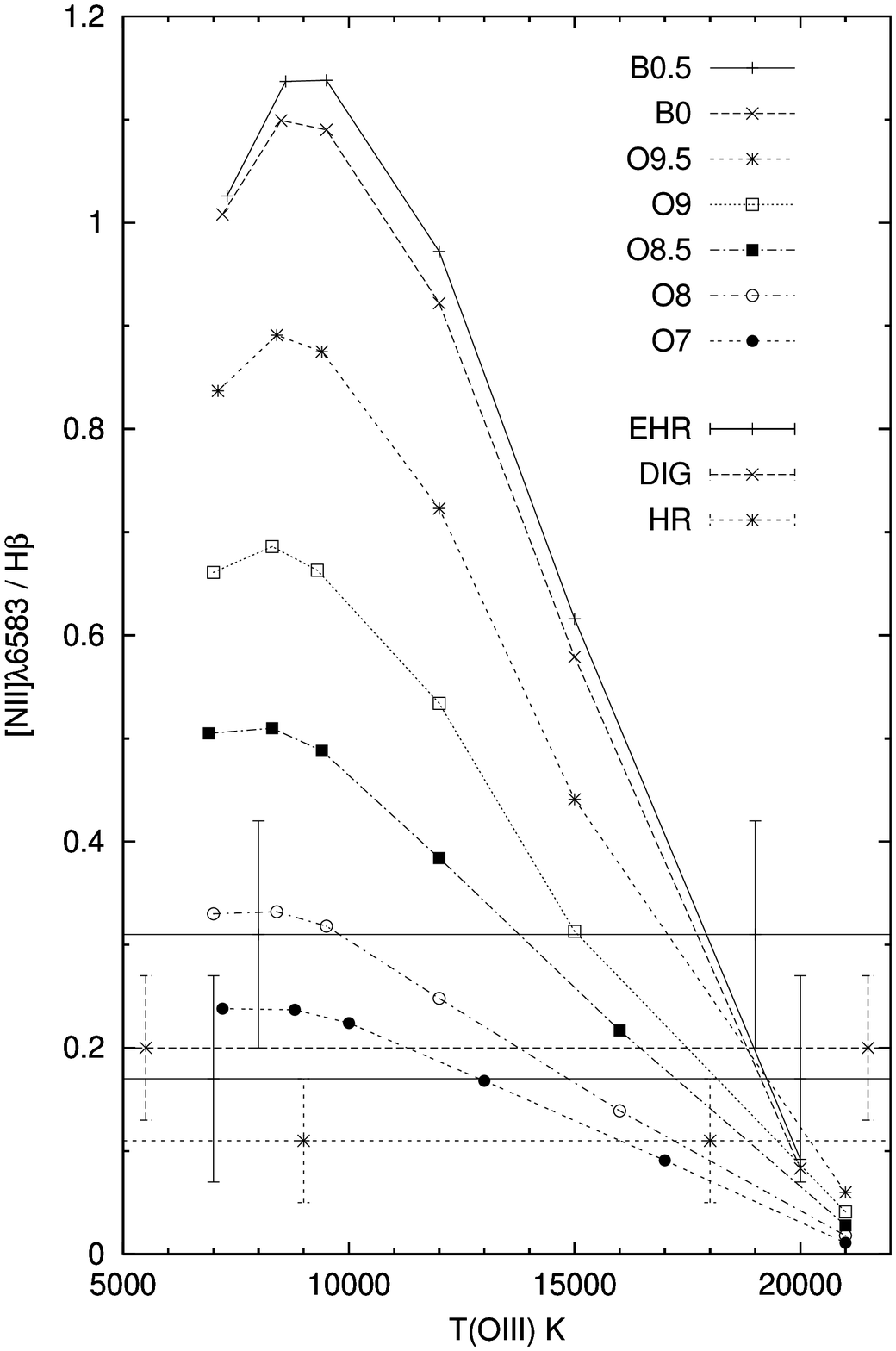}
\includegraphics[width=5.9cm,height=6.8cm,angle=0,clip=t]{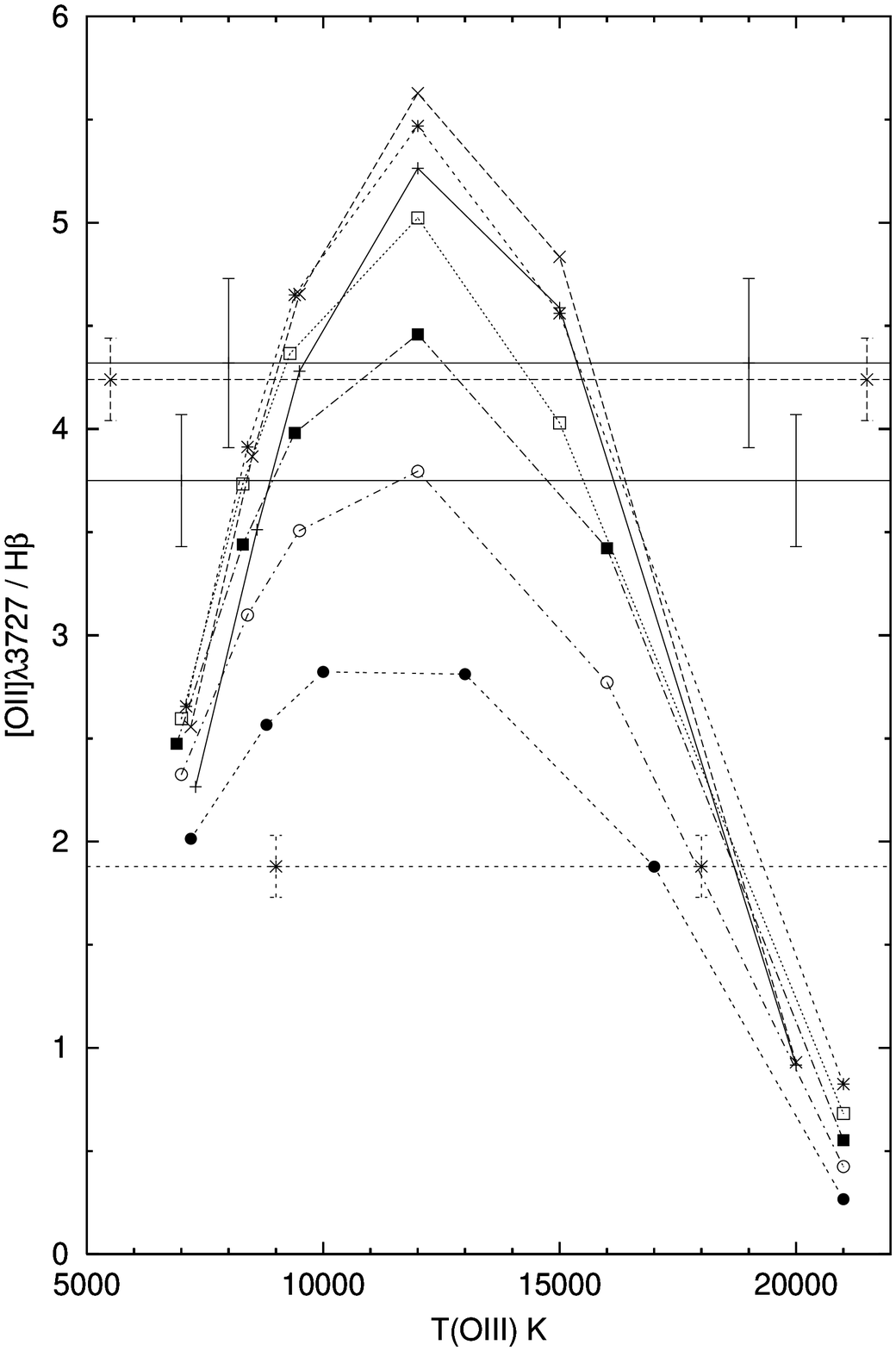}
\includegraphics[width=5.9cm,height=6.8cm,angle=0,clip=t]{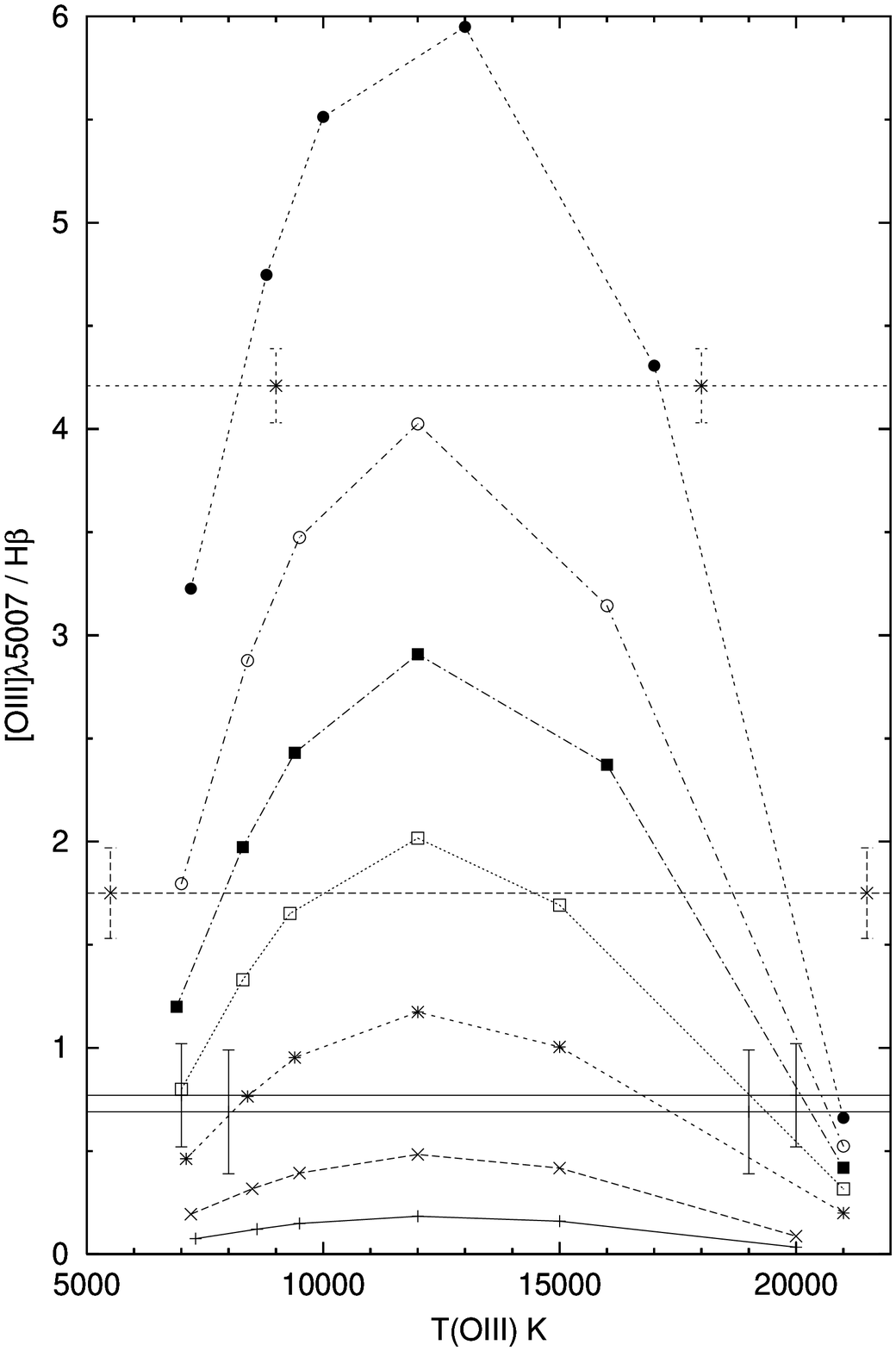}
\caption{Dereddened diagnostic diagrams which constrain the spectral type of the ionizing source and the abundances of the adjacent ISM. 
Horizontal lines, taken from Table~\ref{tab1}, denote averaged line ratios and their error bars for the \ion{H}{ii}-region, the EHRs, and the DIG (see legend). All theoretical line ratios (curved lines with symbols) are plotted as a function O$^{++}$-temperature, because this temperature indicator depends much less on density than the corresponding one for O$^{+}$ or S$^{+}$. Data points from left to right correspond to 1, 2/3, 1/2, 1/4, 1/10, and 1/100 of solar metallicity.}
\label{fig6}
\end{figure*}
\begin{table*}
\caption{Element abundances for the \ion{H}{ii}-region, the EHRs, and the DIG as determined by the direct method (NAT). Values in brackets were derived with the empirical R$_{23}$-calibration. Solar abundances are compiled from the most recent data including Christensen-Dalsgaard (\cite{christensen}) and Grevesse \& Sauval (\cite{grevesse}). The average metallicity $\overline{Z/Z_{\odot}}$ has been calculated from oxygen abundances derived with the NAT and the R$_{23}$-calibration.}
\begin{tabular}{l l l l l l }
\hline
\hline
Parameter & \ion{H}{ii}-region & EHR\_1 & EHR\_2 & DIG & Solar \\
\hline
R$_{23}$ & 0.88 & 0.68 & 0.72 & 0.82 &--\\
$12+\log({\rm He/H})$ & $10.94\pm 0.02$ (10.94) & (10.93) & 10.84 (10.93)& (10.94) & 
10.98\\
$12+\log({\rm O/H})$ & $8.05\pm 0.10$ (8.08) & 7.77 (7.61) & 7.81 (7.68) & 7.91 (7.88)& 
8.71\\
$\log({\rm N/O})$ & $-1.26\pm 0.05$ (-1.30) & $-1.50$ ($-1.54$) & $-1.31$ ($-1.50$)& $-1.33$ ($-1.40$)& $-$0.78\\
$\log({\rm Ne/O})$ & $-0.85\pm 0.10$ & --- & --- & --- & $-$0.71\\
$\log({\rm S/O})$ & $-1.41\pm 0.15$ & $-1.94$& $-1.69 $& $-1.90$ & $-$1.51\\
$\log({\rm O^{+}/O})$ & $-0.731$ & $-0.133$ & $-0.127$ & $-0.268$ & ---\\
$\log({\rm S^{+}/S^{++}})$ & $-$0.722 & --- & --- & --- & ---\\
$\overline{Z/Z_{\odot}}$ & 0.23 & 0.10 & 0.11 & 0.15 & 1.0\\
\hline
\end{tabular}
\label{tab2}
\end{table*}

\subsubsection{The central \ion{H}{ii}-region}
As pointed out above, a comparison of abundances measured within disk and 
extraplanar \ion{H}{ii}-regions can be used to constrain the origin of the latter ones. 
We determined the gas metal content of the central \ion{H}{ii}-region  from the spectrum obtained with slitlet \#10 (Fig.~\ref{fig3}).  

From temperature-sensitive emission-lines, such as [\ion{O}{iii}]$\lambda$4363 
and [\ion{N}{ii}]$\lambda$5755, the electron temperature of the \ion{H}{ii}-region in the 
disk is determined to be $T_{e}=11\,500 \pm 300$\,K. 
The electron density, derived from Fig.~\ref{fig7}, is limited to 
$200\ {\rm cm}^{-3}\le n_{\rm e} \le 1000\ {\rm cm}^{-3}$ 
with the most likely value of $n_{{\rm e}}=300$ cm$^{-3}$. 

Abundances derived using NAT are given in Table~\ref{tab2}, R$_{23}$ abundances are attached in brackets.
The $\log({\rm O/H})$ values derived using these two fully independent methods agree very well. We adopt an averaged oxygen abundance of  $\log({\rm O/H})=-3.94$ $(0.23\,Z_{\odot}$). Observational data for the central \ion{H}{ii}-region
 published by Webster \& Smith (\cite{webster}, region 2), Stasinska et al. (\cite{stasi}), and Zaritsky et al. (\cite{zaritsky}, who used the data of Webster \& Smith (\cite{webster})) amount to $-3.61$ ($0.48\,Z_{\odot}$), $-3.47$ ($0.66\,Z_{\odot}$), and $-3.65$ ($0.44\,Z_{\odot}$), respectively and independently confirm subsolar metallicities for the disk of NGC\,55. 
Compared to the early data obtained with less sensitive equipment our measurement yields by far the lowest oxygen abundance. Data from Stasinska et al. (\cite{stasi}) cover the opposite extreme. However, these variations can be explained by different slit positions. 
On the VLT H$\alpha$-image the so called ``central'' \ion{H}{ii}-region actually consists of three blobs of emission within a common halo. Our slit cuts through  the southeastern one which has apparently the highest temperature and hence the lowest oxygen abundance.

The reference disk abundance is finally determined by simply averaging our and published data for the same region (region ``2'' in Webster \& Smith (\cite{webster}) and ``central'' in Stasinska et al. (\cite{stasi})). This yields $-3.64$ and translates into 45$\%$ of solar metallicity.
Due to uncertainties of about 0.1\,dex the element oxygen abundance is expected to range between $0.36\,Z_{\odot}$ and $0.56\,Z_{\odot}$.

As can be seen from Fig.~\ref{fig6}, the ionizing sources within the central \ion{H}{ii}-region can be approximated with CLOUDY by an averaged spectrum typical for an O7-star. At temperatures of 11\,500\,K, solar abundances should be scaled by factors varying from 0.25 to 0.5. This independently confirms the results obtained with the NAT and the R$_{23}$-method.   

Since Ferguson et al. (\cite{fergi}) calculated the H$\alpha$-luminosity of this region to be $L_{{\rm H}\alpha}= 9 \times 10^{39}\,{\rm erg\ s}^{-1}$, we
 can also estimate the number of ionizing stars via Eq.~(\ref{eq1}). If all stars within the disk \ion{H}{ii}-region were of type O7, the number of ionizing sources would amount to 3 clusters containing a total of 290 stars. 

\subsubsection{Extraplanar \ion{H}{ii}-regions}
\begin{figure}
\centering
\includegraphics[width=5.5cm,height=8.75cm,angle=-90]{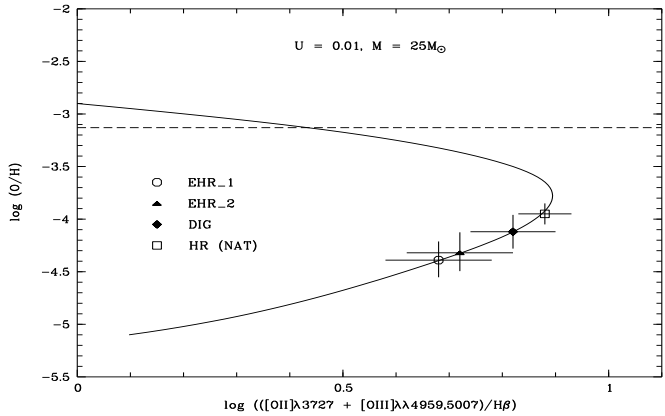}
\caption{Oxygen abundances based on the R$_{23}$-calibration. The solid line 
represents a CLOUDY model (McGaugh \cite{mcg91}) which adopts an ionization 
parameter $U$ of 0.01 and a stellar mass of 25\,M$_{\odot}$ (representing an O9 star of temperature 42\,000\,K with a metallicity of $0.1\,Z_{\odot}$ (Maeder \cite{maeder})). For comparison, the solar oxygen abundance (dashed line) is also plotted.} 
\label{fig8}
\end{figure}

Abundances for EHR\_1 and EHR\_2, listed in Table~\ref{tab2}, are calculated using again the R$_{23}$ and the NAT-method.
The $\mathrm{R}_{23}$-calibration used is shown in Fig.~\ref{fig8} for a stellar spectrum of type O9 with an ionization parameter of $U=0.01$. This plot suggests that both EHRs have $\log(\mathrm{O/H})$ abundances which are much lower than those for the central disk \ion{H}{ii}-region, namely 8$\%$ and 9$\%$ of the solar value. Compared to the reference abundance of the disk ($45\%\,Z_{\odot}$), their metal content appears to be reduced by about a factor of 5 by the $\mathrm{R}_{23}$-method alone. 

Since [\ion{O}{iii}]$\lambda$4363 is too weak to allow a direct measurement of the temperature, assumptions on the gas temperatures
are required as input for NAT. We adopted the value for the central \ion{H}{ii}-region (11\,500\,K) to be representative for both EHRs. This yielded $\log(\mathrm{O/H})$ abundances in close accordance (within  0.1\,dex) with those from the $\mathrm{R}_{23}$-method.  
Significantly higher temperatures would give even lower abundances and should push [\ion{O}{iii}]$\lambda$4363 above the detection limit. Temperatures below 9600\,K can be ruled out for both EHRs, because they indicate abundances equal to or larger than those for the disk. 

According to Fig.~\ref{fig6}, EHR\_1 seems to be ionized by an O9 or O9.5 star, whereas EHR\_2 is excited by a stellar source of spectral type B0 or O9.5. This is consistent with our CLOUDY simulations presented in Sect.~3.1. As electron temperatures predicted by these models vary from 13\,600\,K to 15\,300\,K, abundances derived with CLOUDY (Fig.~\ref{fig6}) indicate values between 0.1 -- 0.25 of solar metallicity. This is in good agreement with the findings from the $\mathrm{R}_{23}$-calibration and the NAT calculation.

\subsubsection{Diffuse Ionized Gas}
The diffuse ionized gas observed in galaxies is presumably blown out into the halo by the OB winds and SNe that occurred in star-forming regions in the disk.  
Typical adopted temperatures for a number of galaxies studied are around 10\,000\,K and densities vary between 0.1 and 0.01\,cm$^{-3}$ (for a review see Dettmar \cite{dettmar}). One issue which still remains to be unexplained concerns the ionization mechanism of the DIG. Pure photoionization by OB stars 
(e.g., Domg\"orgen \& Mathis \cite{doma}), as well as a combination of different heat sources have been proposed, ranging from passing shock fronts (Shull \& McKee \cite{shull}; Rand \cite{rand}),  turbulent mixing layers (Slavin et al. \cite{slavin}), magnetic reconnection (Lesch \& Bender \cite{lesch}; T\"ullmann et al. \cite{me}) to non-linear Landau damping (Lerche \& Schlickeiser \cite{lerche}).

Estimates of the energy required to ionize the DIG of the Milky Way Galaxy 
(Reynolds \cite{rey}) clearly favour pure photoionization by O stars. For the following abundance determination we also assume that this mechanism is responsible for the bulk of the DIG emission in NGC\,55.  The DIG covered by slitlet \#13 (Fig.~\ref{fig3}) apparently receives a substantial amount of ionizing radiation from the extended underlying star-forming region. 
In principle radiative transfer, ionization equilibrium and thermal equilibrium of such a geometry is insufficiently described by standard photoionization models (and therefore also by the standard methods to estimate chemical abundances), which assume a central point source embedded in a spherical geometry. Although we have developed methods to treat the radiative transfer of such geometries more accurately (T\"ullmann \cite{me2}), sophisticated numerical models which estimate chemical abundances similar to NAT, $\mathrm{R}_{23}$ or the CLOUDY model grids are not yet available. We therefore resort to those three methods.

Here we present the first estimation of chemical abundances of the DIG in NGC\,55. 
Since also for the DIG reliable temperature estimators are lacking, we assume a temperature of 13\,000\,K. As we will see this is the lowest temperature which keeps the [O/H] value derived for the DIG in the range covered by the two EHRs. Lower temperatures and therefore significantly higher metallicities for the DIG seem unreasonable. On the other hand, a higher temperature for the DIG in comparison to the EHRs at the same metallicity can be justified by lower densities, at which collisional cooling is less effective and recombination times are longer.
Densities based upon the [\ion{S}{ii}]$\lambda$6717,6731 emission-lines (Osterbrock \cite{oster}) fall within the low density limit and are therefore $\le 10\ {\rm cm}^{-3}$. Although X-ray and DIG emission are well aligned along the filament, it would be interesting to investigate if DIG of rather low densities actually represents the wall of a chimney.

Fig.~\ref{fig6} reveals that DIG at the adopted temperature is likely ionized by stars of type O7 -- O9. In case the OB stars of the central \ion{H}{ii}-region are responsible for the ionization of the DIG, a dilute and slightly softer radiation field, which is similar to that of the disk (average O7 type spectrum), is expected.

In comparison to the EHRs, the harder photon field ionizing the DIG is also consistent with emission-line data presented in Table~\ref{tab1} and Fig.~\ref{fig4}. 
Element abundances for the DIG vary between 0.1 and $0.25\,Z_{\odot}$. The 
corresponding quantities determined by NAT and R$_{23}$ are limited to $0.15\,Z_{\odot}$ (cf. Table~\ref{tab2}). 

Relative to the disk \ion{H}{ii}-region the DIG certainly has a lower element abundance (a rather low electron temperature of about 10\,000\,K would be required to bring the [O/H] ratio of the DIG measurement by NAT up to the [O/H] abundance characteristic of the disk). 
In the framework of a large-scale transport of matter into the halo and back 
(Shapiro \& Field \cite{shapi}; Norman \& Ikeuchi \cite{norman}), it appears likely that the gas processed in star formation has been driven out of the disk plane. Since the extraplanar DIG is no longer involved in star formation, lower element abundances are a natural consequence.

\begin{figure}
\centering
\includegraphics[width=5.5cm,height=8.75cm,angle=-90]{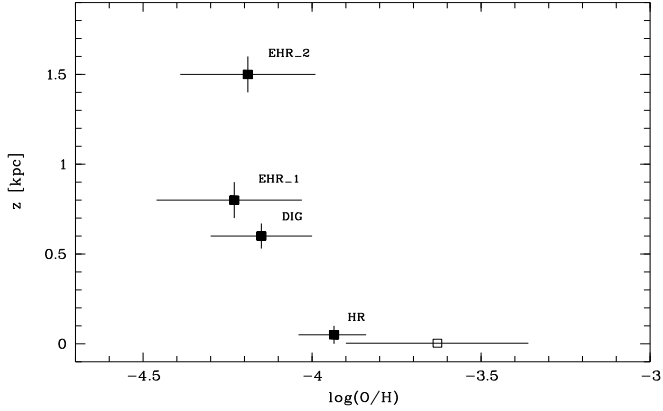}
\caption{Oxygen abundances along the minor axis of NGC\,55. The open symbol represents averaged data obtained for the disk (see text), whereas the error bar represents the scatter among these values. Filled squares address our data. 
The datapoint labeled ``HR'' has been slightly shifted along $z$ to separate error bars.}
\label{fig9}
\end{figure}

If for all relevant objects $\log ({\rm O/H})$ as a tracer of the gas metal content (cf. Table~\ref{tab2}) is plotted vs. $|z|$, their distance above the disk plane, the strong decrease in the gas metal content towards the halo can be seen much more clearly (Fig.~\ref{fig9}). 
Data for the disk measured at different radial distances from the center has been compiled again from Webster \& Smith (\cite{webster}), Stasinska et al. (\cite{stasi}), and Zaritsky et al. (\cite{zaritsky}) and averaged to give a single reference point.
Our value of the central oxygen abundance is consistent within errors with the averaged disk abundance.
  
Differences in $\log ({\rm O/H})$ amount to 0.6\,dex up to $z=1.5$\,kpc and indicate that oxygen is less abundant in the halo by about a factor of 4.

\subsection{The origin of the extraplanar \ion{H}{ii}-regions}
In order to constrain the positional origin of the molecular matter out of which stars formed that ionized the clouds, one can estimate the maximum path length which stars and clouds could travel during the lifetime of the embedded stars. The radial velocities observed for the two EHRs are within their errors compatible with the velocities observed in \ion{H}{ii}-regions in the main body of NGC\,55 (Table~\ref{tab1}). Proper motions are not known, but since both EHRs are located relatively high above the disk plane, these velocities could be quite high.  Disregarding for a moment the hydrodynamical effects expected when gas moves through gas, one can put an absolute upper limit on the linear proper motion using a rather high escape velocity of 200 km\,s$^{-1}$ (Leonard 1991). In addition, one can use a lower limit of the spectral type of the most massive stars which are still able to ionize the \ion{H}{ii}-regions. Mean lifetimes for such stars of type B0\,V are approximately $7.5\times 10^{6}$\,yrs. 

With these parameters, we calculate an upper limit for the linear distance of about 1.5\,kpc. Hence, {\it potentially} both EHRs {\it could} have originated in the disk and then been expelled by some unknown mechanism at the escape velocity. 

This hypothesis is, however, quickly ruled out if we consider the enormous drag, the gas phase of these nebulae encountered on its way out of the disk into the halo. It is therefore much more natural to assume that the molecular clouds of both star-forming regions had been present at or near their present locations high above the disk long before current star formation occurred. Two questions emanate from this hypothesis: (a) how did matter get there in a quantity to form neutral, dense cores from which stars could form, and (b) what triggered the collapse to form those stars?
  
The H$\alpha$-image provides some clues. Both regions are within the greater area of outflow activity connected to the rather intense star formation activity in the central part of NGC\,55. Both regions are located near grand arcs of ionized gas that are likely indicators of supershells created by the snowplow effect of processed matter  flowing out into the halo. One putative scenario might be that the natal clouds are remnants from similar outflows during an early epoch of central star formation, meanwhile cooled down and condensed. During the supershell phase of the current star formation activity in NGC\,55, these molecular clouds were hit by the compression wave and star formation was triggered in the EHRs a few million years ago. 

In that picture, the EHRs are just another in situ tracer of the long-range effects which ionization and stellar winds from the central star-forming region can have on the gas phase high above the disk. Yet they are also tracers of star formation-enriched material brought high above the plane in thin, ionized form, to cool down, form molecular clouds, and be triggered to form stars when the next major star formation event hits the underlying galactic disk. 

\section{Summary and Conclusions}
The morphology of the continuum (Fig.~\ref{fig4}) and the nebular emission-line characteristics (Fig.~\ref{fig5}) of both extraplanar objects confirmed our initial assumption of low-excitation \ion{H}{ii}-regions in the halo of NGC\,55.  Final evidence of the \ion{H}{ii}-region character was given by CLOUDY simulations which could reproduce the observed spectra reasonably well and also constrain the spectral type of the ionizing sources.

EHR\_1, located 0.8\,kpc in the northern halo, has a diameter of about 17\,pc, a gas density of 600\,$\mathrm{cm}^{-3}$, and is most likely ionized by O9.5 stars. 
For EHR\_2 in the southern halo, the distance to the disk plane has been determined to be 1.5\,kpc. As the H$\alpha$-luminosity of this region is known, we could restrict the number of ionizing sources to a single star, either of type B0 or O9.5. The resulting diameter of the Str\"omgren sphere amounts to 22\,pc. Densities are about 100\ $\mathrm{cm}^{-3}$. For both EHRs the gas temperature is of the order of 11\,500\,K.

In comparison to the central \ion{H}{ii}-region and the DIG, the ionizing radiation field of the EHRs is actually the softest. 
DIG in the disk-halo interface seems to be ionized by an average spectrum typical for O8 type stars. Gas temperatures are most likely around 13\,000\,K. The number of ionizing stars within the \ion{H}{ii}-region of the disk has been estimated to be 290. They are apparently of type O7 or younger. Averaged gas temperatures are about 11\,500\,K and densities range between 200\,cm$^{-3}$ and 1000\,cm$^{-3}$, with 300\,cm$^{-3}$ being the most likely value.

After the \ion{H}{ii}-region character has been established for both EHRs, the question of star formation in galaxy halos had to be answered. This was done by measuring element abundances for the extraplanar objects and comparing them to those determined for the central disk \ion{H}{ii}-region. 
Three different abundance estimators that produced fully consistent results have been used. 
We determined the metal abundances for the central \ion{H}{ii}-region, the EHRs, and the DIG, to be 23$\%$, 10$\%$, and 15$\%$ of solar metallicity, respectively. 
With abundances derived for the disk and the halo gas, we could establish a strong jump in abundances, which implies that oxygen is less abundant in the halo by about a factor of 4.

The differences between the metallicities of the central \ion{H}{ii}-region and the EHRs are significant, but the actual value is strongly dependent upon the assumptions made about the electron temperatures in the EHRs. The metallicity of the DIG is certainly lower than that of the disk, but to determine if it is different from that of the EHRs, much deeper spectra are required.

Since travel time estimations are not applicable here, both EHRs must have 
originated within the halo of NGC\,55, far away from traditional star-forming sites.

Both regions are located above the main star-forming complex. This immediately raises the question how molecular gas clouds can sustain the energetic UV radiation from the center, cool, collapse and finally form stars.  
The most simple and natural explanation to this question is, that these regions were present and collapsed before the current burst of star formation occurred.

For future work, it would be useful to test the ESF scenario for a larger sample of galaxies, to investigate initial formation conditions for EHRs, and to check if the central stars can separate within their lifetime far enough from their gaseous environment to contribute to the observed stellar halo population.

\begin{acknowledgements}
RT received financial support by the GRK\,787 ``Galaxy Groups'' and the DLR through grant 50OR0102. The research of AMNF has been supported by a Marie Curie Fellowship of the European Community under contract number HPMF-CT-2002-01758. 
Norbert Rainer from ESO provided important help with the FORS-pipeline.
\end{acknowledgements}

\end{document}